\documentclass[preprint,prd,aps,nofootinbib]{revtex4}
\usepackage{amsmath}
\usepackage{psfig}

\begin{document}                                                 
\newcommand{\be}{\begin{equation}}
\newcommand{\ee}{\end{equation}}
\newcommand{\ba}{\begin{eqnarray}}
\newcommand{\ea}{\end{eqnarray}}
\newcommand{\bc}{\begin{center}}
\newcommand{\ec}{\end{center}}
\newcommand{\vs}{\vspace*{3mm}}
\newcommand{\dis}{\displaystyle}
\newcommand{\bay}{\begin{array}}
\newcommand{\eay}{\end{array}}
\def\RN{Reis\-sner-Nord\-str\"{o}m }
\def\rc{\rho_{\rm crit}}
\def\rl{\rho_\Lambda}
\def\rt{\rho_{\rm tot}}
\def\ie{{\it i.e.\;}}
\def\lp{\ell_{\rm Pl}}
\def\mp{m_{\rm Pl}}
\def\tp{t_{\rm Pl}}
\def\tf{t_{\rm FP}}
\def\om{\Omega_{\rm M}}
\def\oa{\Omega_{\Lambda}}
\def\omz{\Omega_{\rm M0}}
\def\oaz{\Omega_{\Lambda 0}}
\def\ot{\Omega_{\rm tot}}
\def\tla{\widetilde{\lambda}_\ast}
\def\tom{\widetilde{\omega}_\ast}
\def\luv{\lambda_\ast^{\rm UV}}
\def\guv{g_\ast^{\rm UV}}
\def\lir{\lambda_\ast^{\rm IR}}
\def\gir{g_\ast^{\rm IR}}
\def\lir{\lambda_\ast^{\rm IR}}
\def\gir{g_\ast^{\rm IR}}
\renewcommand{\baselinestretch}{1.5}

\begin{flushright}
MZ-TH/03-04\\
\end{flushright}

\title{Confronting the IR Fixed Point Cosmology with \\
High Redshift Observations}
\author{Eloisa Bentivegna}
\email{ebe@ct.astro.it}
\affiliation{Dipartimento di Fisica e Astronomia, 
Universit\`a degli Studi di Catania, Via S.Sofia 64, I-95123 Catania, Italy}
\author{Alfio Bonanno}
\email{abo@ct.astro.it}
\affiliation{INAF, Osservatorio Astrofisico di Catania, Via S.Sofia 78,
I-95123 Catania, Italy}
\altaffiliation[Also at ]{INFN, Sezione di Catania, Via S.Sofia 73}
\author{Martin Reuter}
\email{reuter@thep.physik.uni-mainz.de}
\affiliation{Institut f\"ur Physik, Universit\"at Mainz\\
Staudingerweg 7, D-55099 Mainz, Germany}
\renewcommand{\baselinestretch}{1}
\begin{abstract}
We use high-redshift type Ia supernova and compact radio source data in order to test the infrared (IR) fixed 
point model of the late Universe which was proposed recently. It describes a cosmology 
with a time dependent cosmological constant and Newton constant whose dynamics arises from an 
underlying renormalization group flow near an IR-attractive fixed point. Without any finetuning 
or quintessence field it yields $\Omega_{\rm M}=\Omega_{\Lambda}=1/2$. Its characteristic $t^{4/3}$-
dependence of the scale factor leads to a distance-redshift relation whose predictions are 
compared both to the supernova and to the radio source data. According to the $\chi^2$ test, 
the fixed point model reproduces 
the data at least as well as the best-fit (Friedmann-Robertson-Walker) standard cosmology. 
Furthermore, we extend the original fixed point model by assuming that the fixed point 
epoch is preceded by an era with constant $G$ and $\Lambda$. By means of a Monte Carlo simulation 
we show that the data expected from the forthcoming SNAP satellite mission could detect the transition 
to the fixed point regime provided it took place at a redshift of less than about $0.5$.
\end{abstract}
\renewcommand{\baselinestretch}{1.5}
\maketitle

\section{Introduction}
In recent years renormalization group (RG) techniques have been extensively applied 
in cosmology, for instance in a classical averaging scenario \cite{classav} or in discussing
the late-time behavior of the classical Einstein equations \cite{bel}. 
As for possible quantum gravitational effects, an exact functional RG equation \cite{ber} for 
Quantum Einstein Gravity (QEG) has been introduced \cite{mr,geo} and, within certain approximations 
(truncations of ``theory space''), its predictions for the scale dependence of Newton's constant 
$G$ and the cosmological constant $\Lambda$ have been worked out \cite{ol,ol2,frank,per}.

In particular, it was found \cite{ol,souma,max} that the ultraviolet (UV) behavior of QEG is 
likely to be governed 
by a non-Gaussian RG fixed point which would render the theory nonperturbatively renormalizable. 
If so, QEG is mathematically consistent and predictive at arbitrarily short distances and, in cosmology, 
at times arbitrarily close to the initial singularity. In ref. \cite{br1} we investigated how the standard 
Friedmann-Robertson-Walker (FRW) cosmology gets modified when the scale dependence of $G$ and $\Lambda$ 
is taken into account. We ``RG improved'' Einstein's equation by replacing $G \to G(k)$, 
$\Lambda \to \Lambda(k)$, where $k$ is the running mass scale which may be identified with 
the inverse of the cosmological time in a homogeneous and isotropic Universe, $k \propto 1/t$. 
In this manner the RG running gives rise to a dynamically evolving, time dependent $G$ and $\Lambda$ \cite{cw}. 
The improvement of Einstein's equation can be based upon any RG trajectory $k \mapsto (G(k), \Lambda(k))$ 
obtained as an (approximate) solution to the exact RG equation of QEG. In particular, if the non-Gaussian 
UV fixed point predicted by all known solutions does indeed exist, $G(k)$ and $\Lambda(k)$ scale in a very 
simple manner as $k \to \infty$. In fact, the dimensionless Newton constant $g(k) \equiv k^2 G(k)$ and 
cosmological constant $\lambda(k) \equiv \Lambda(k)/k^2$ are attracted towards their fixed point values $g_\ast$ 
and $\lambda_\ast$, respectively, and therefore, in the fixed point regime,
\begin{subequations}
\label{eq:traj}
\ba
G(k)=\frac{g_\ast}{k^2}\\
\Lambda(k)=\lambda_\ast \, k^2
\ea
\end{subequations}
In ref. \cite{br1} the improved cosmology resulting from the trajectory (\ref{eq:traj}) was analyzed in 
detail. As the fixed point ($g_\ast, \lambda_\ast$) is approached for $k \to \infty$, it 
applies to the very early Universe ($t \to 0$). It turned out that the cosmology of this ``Planck 
era'' is described by an essentially unique attractor solution to the cosmological evolution 
equations which might provide a solution to the horizon and flatness problems of standard cosmology 
without invoking an inflationary era\footnote{A similar RG improvement of black hole spacetimes 
can be found in refs. \cite{bh2,bh1}.}. 

One of the remarkable properties of the attractor solution is that it dynamically adjusts the vacuum 
energy density $\rl \equiv \Lambda/8 \pi G$ so as to equal precisely the matter energy density. In 
units of the critical density, the spatially flat solution has
\be
\oa(t) = \om(t) = \frac{1}{2}
\ee
at any time during the fixed point era. This adjustment mechanism has led to the speculation \cite{br2} 
that not only the very early, but also the very {\it late} history of the Universe is described by a 
RG-improved Einstein equation based upon the trajectory (\ref{eq:traj}). In this scenario one postulates 
that, in addition to the UV-attractive fixed point discussed so far, there exists a second fixed point 
($\gir, \lir$) in ($g, \lambda$)-space towards which every trajectory $k \mapsto (g(k), \lambda(k))$ 
within a certain basin of attraction is attracted for $k \to 0$. On very large scales the cutoff 
identification $k \propto 1/t$ should still be correct so that for $t \to \infty$ the IR fixed point 
determines the asymptotically late cosmology. The IR fixed point hypothesis implies that for $k \to 0$ 
or $t \to \infty$, too, $G$ and $\Lambda$ evolve accordingly to (\ref{eq:traj}). (In the sequel we shall 
omit the superscripts ``IR'' from $g_\ast$ and $\lambda_\ast$ but it should be kept in mind that the 
values of $g_\ast$ and $\lambda_\ast$ are different at the two fixed points.) The main motivation for 
this hypothesis comes from the fact that, provided the Universe is spatially flat and has entered the 
fixed point regime already, the densities $\oa$ and $\om$ are unambiguously predicted to assume the 
value $0.5$. This is intriguingly close to the values $\oa \approx 0.7$, $\om \approx 0.3$ favored 
by recent observations when interpreted within standard FRW cosmology. Hence the IR fixed point might 
provide a natural solution to the ``cosmic coincidence problem'' \cite{coscon2} which does not require 
quintessence field. 

As for their actual verification within QEG, the UV and the IR fixed points have a rather different 
status for the time being. While the existence of the UV fixed point has actually been demonstrated 
within certain approximations whose reliability was tested in detail by means of rather involved 
calculations \cite{ol,ol2}, to date there is no comparable evidence for the IR fixed point. (See refs. 
\cite{frank,tsamis}, however.) 
The technical reason is that it is extremely difficult to follow the RG flow from the UV all the way 
down to the IR. Therefore a $k \to 0$ RG-running of the form (\ref{eq:traj}) has the status of 
a conjecture for the time being; it is motivated by its remarkable phenomenological success. 
We also emphasize that the RG-running of $G$ and $\Lambda$ should not necessarily be thought of as 
arising from quantum effects; in principle it could also result from a purely classical averaging 
(coarse graining) \cite{classav}.

The general properties of the IR fixed point (IRFP) cosmology were first discussed in ref. \cite{br2} and will 
be reviewed briefly in subsection \ref{sec:review}. In ref. \cite{pert} the theory of small perturbations about 
the IRFP cosmology has been developed and its consequences for the cosmological structure formation 
were discussed.

In the present paper we are going to test the IRFP cosmology by confronting its predictions with 
the currently available data on distant type Ia supernovae and compact ($\sim$ milliarcsecond) radio sources. 
The relevant quantities which can be compared 
to the data are the luminosity distance $d_L(z)$ and the angular diameter distance $d_A(z)$, as functions of the 
redshift. In standard FRW 
cosmology these functions depend on $\om$ and $\oa$, and these parameters are fitted so as to reproduce 
the observations as well as possible. In the IRFP cosmology instead, the functions $d_L(z)$ and $d_A(z)$ contain 
{\it no free parameters} at all. Hence an acceptable fit to the data would be a rather 
non-trivial success of the IRFP scenario.

In particular we apply the $\chi^2$ test to the
combined dataset of 92 type Ia supernovae discovered by the Supernova Cosmology Project (SCP)
\cite{perl} and the High-z Supernova Search Team (HzST) \cite{riess} and to the radio source
data set of \cite{gurv}.
As far as the $\chi^2$ test is concerned, we shall show that the IRFP cosmology 
has a reduced $\chi^2$ value which is basically the same as that of the best-fit FRW solution.
In the Appendix, we also apply the median statistics analysis \cite{gott,avel} to the SNe Ia fit, 
and then use the Bayesian 
approach to test the robustness of the hypothesis that the correct model of the late Universe 
is the IRFP cosmology,  and we show that the posterior probability that the late
Universe is described by the IRFP cosmology is $36\%$. The corresponding probability for the 
best-fit FRW cosmology is only $24\%$. These numbers are obtained from a specific choice of the priors
which was used in the literature already earlier, but of course, as always, it contains a certain
degree of arbitrariness.

The rest of this paper is organized as follows. In Section II we discuss the 
basic properties of the IRFP cosmology and introduce a possible extension of the model
which describes the complete transition from a preceding FRW era to the IRFP cosmology.
Section III is devoted to the calculation of the luminosity distance, and to the 
data analysis by means of the $\chi^2$ test.  
In Section IV we perform an analogous cosmological test, based on the angular 
diameter distance-redshift relation.
In Section V we discuss arbitrary power law cosmologies with variable $G$ and $\Lambda$ 
which generalize the IRFP model. In Section VI we derive a constraint on the IRFP from
the CMBR acoustic peaks, and Section VII is devoted to the conclusions.
\section{The model}\label{sec:model}
\subsection{The fixed point cosmology}\label{sec:review}
In this subsection we shall review the basic properties of the IRFP cosmology. 
It assumes spacetime to be homogeneous and isotropic so that it can be 
described by a standard Robertson-Walker metric containing the scale factor 
$a(t)$ and the parameter $K=0, \pm 1$ which distinguishes the three types of 
maximally symmetric 3-spaces of constant cosmological time $t$. The time evolution 
is governed by Einstein's equation $R_{\mu\nu} - \frac{1}{2} g_{\mu\nu} R = -\Lambda g_{\mu\nu} + 
8 \pi G T_{\mu\nu}$ with the conserved energy-momentum tensor $T_\mu^{\phantom{\mu}\nu}$ = diag 
($-\rho, p, p, p$). The equation of state is assumed to be of the form $p(t)=w \rho(t)$, 
with $w > -1$ an arbitrary constant. Now we ``RG improve'' Einstein's equation by 
replacing $G \to G(t)$ and $\Lambda \to \Lambda(t)$ which leads to the following coupled system 
of evolution equations \cite{br1,br2}:
%
\begin{subequations}
\label{eq:system}
\ba
\label{eq:IFE}
&&\left(\frac{\dot{a}}{a}\right)^2+\frac{K}{a^2}=\frac{1}{3}\Lambda+
\frac{8\pi}{3}  G \rho\\
\label{eq:tensor}
&&\dot{\rho}+3(1+w)\frac{\dot{a}}{a}\rho=0\\
\label{eq:BI}
&&\dot{\Lambda}+8 \pi \rho\; \dot{G}=0\\
\label{eq:RGflow}
&&G(t) \equiv G(k=k(t)), \quad \Lambda(t) \equiv \Lambda(k=k(t))
\ea
\end{subequations}
Eq. (\ref{eq:IFE}) is the familiar Friedmann equation with a time dependent 
$G$ and $\Lambda$, and eq. (\ref{eq:tensor}) expresses the conservation of 
$T_{\mu\nu}$. Eq. (\ref{eq:BI}) is a novel integrability condition which ensures 
that the RHS of Einstein's equation has vanishing covariant divergence.

The equations (\ref{eq:RGflow}) express the fact that the {\it time} dependence 
of $G$ and $\Lambda$ is the consequence of a more fundamental {\it scale} dependence 
of these quantities. We describe physics at a typical distance scale $\ell \equiv k^{-1}$ 
by means of a scale dependent gravitational action $\Gamma_k[g_{\mu\nu}]$ which 
should be thought of as a coarse grained free energy functional in the sense of Wilson. 
It encapsulates the effect of all metric fluctuations with momenta larger than the 
IR cutoff scale $k$, while those with smaller momenta are not yet ``integrated out''. 
In very general terms, $\Gamma_k$ describes the dynamics of fields ``averaged'' over 
spacetime volumes of linear extension $k^{-1}$. The $k$-dependence of $\Gamma_k$ is 
governed by an exact RG equation whose precise nature is not important here. The 
quantum mechanical flow equation for the exact average action of QEG \cite{mr} would be one 
example, the classical equations of ref. \cite{classav} are another. Generically $\Gamma_k$ is an 
arbitrary diffeomorphism-invariant functional depending on infinitely many 
generalized coupling constants which multiply all possible invariants which can be 
constructed from $g_{\mu\nu}$. We assume that we are in a regime where only the 
Einstein-Hilbert invariants $\int d^4x \sqrt{-g} \, R$ and $\int d^4x \sqrt{-g}$ are 
important so that the only running couplings retained are their coefficients 
$G(k)$ and $\Lambda(k)$. If $\Gamma_k$ is of the Einstein-Hilbert form, the 
effective, i.e. scale dependent field equations look like the conventional 
Einstein equation with the replacement $G \to G(k)$, $\Lambda \to \Lambda(k)$, 
where $k \mapsto (G(k), \Lambda(k))$ is an approximate solution to the RG equation.

The final step of the improvement program consists in converting the $k$-dependence 
of $G$ and $\Lambda$ to a time dependence. In ref. \cite{br1} we discussed in detail that 
in a Robertson-Walker spacetime, to leading order, the IR cutoff should be 
identified as
\be
k(t)=\frac{\xi}{t}
\ee
where $\xi$ is a positive constant.

Before discussing the solutions of the system (\ref{eq:system}) some definitions are convenient.
We define the vacuum energy density $\rl$, the total energy density
$\rt$ and the critical energy density $\rc$ according to
\ba
&&\rho_{\Lambda}(t) \equiv \frac{\Lambda(t)}{8 \pi G(t)}, \qquad
\rho_{\rm{tot}}(t) \equiv \rho + \rl,\\
\label{eq:cdens}
&&\rho_{\rm{crit}}(t) \equiv \frac{3} {8 \pi G(t)} \left(\frac{\dot{a}}{a}
\right)^2
\ea
with $H \equiv \dot a / a$. Hence we may rewrite the improved Friedmann equation (\ref{eq:IFE}) in the form
\be
\label{eq:IFE2}
\frac{\dot a^2+K}{a^2} = \frac{8 \pi}{3} G(t) \rt
\ee
Frequently we refer the various energy densities to the critical density
(\ref{eq:cdens}):
\be
\Omega_{\rm M} \equiv \frac{\rho}{\rho_{\rm crit}}, \quad \Omega_{\Lambda} \equiv 
\frac{\rho_{\Lambda}}{\rho_{\rm crit}}, \quad \Omega_{\rm tot} \equiv  \Omega_{\rm M} + 
\Omega_{\Lambda} \equiv \frac{\rho_{\rm tot}}{\rho_{\rm crit}}
\ee
Another convenient abbreviation is $\Omega_{\rm K} \equiv 1-\ot=1-\om-\oa$.
It follows from these definitions that
\be
\label{eq:lambda}
\Lambda(t)=3\;\Omega_{\Lambda}(t)H^2(t)
\ee
and the Friedmann equation (\ref{eq:IFE2}) becomes
\be
\label{eq:kol}
K=\dot a^2 \; [\Omega_{\rm tot}-1]=-\dot a^2 \; \Omega_K
\ee
For a spatially flat universe ($K=0$) we need $\rt=\rc$, as in
standard cosmology. In this case the definition (\ref{eq:cdens}) entails
\be
\label{eq:costante}
\rt(t) \; G(t) \; H^2(t)=\frac{3}{8 \pi} \qquad (K=0)
\ee

It is important to recall from \cite{br2} that the standard experimental value of
Newton's constant, $G_{\rm{exp}}$, does not coincide with the value $G(k=\xi/t_0)$
which is relevant for cosmology today, i.e. for $t=t_0$.
In fact, $G_{\rm{exp}}$ is measured (today) at $k_{\rm{exp}}=\xi/\ell$, 
where $\ell$ is a
typical laboratory or solar system length scale. Thus, in terms of the running
Newton constant, $G_{\rm{exp}}=G(k=k_{\rm{exp}})$, since $1/\ell \gg 1/t_0$, and since in
presence of several scales the relevant cutoff is always the larger one. On the
basis of the antiscreening character of gravity \cite{mr}  we might expect that 
$G(t_0) > G_{\rm{exp}}$ therefore.

Note that the critical density 
(\ref{eq:cdens}) is defined in terms of the {\it cosmological}
Newton constant $G(t)$. In particular in order to parametrize matter 
densities $\rho$ obtained by non-cosmological measurements, it is sometimes 
more convenient to refer $\rho$ to a modified ``critical'' density defined in 
terms of the {\it experimental} value $G_{\rm{exp}}$:
\be
\rc^{\rm{exp}}(t) \equiv \frac{3H^2(t)}{8 \pi G_{\rm{exp}}}
\ee
A spatially flat universe requires $\rt=\rc$, but $\rt$
might well be different from $\rc^{\rm{exp}}$. For the relative matter 
density referring to $\rc^{\rm{exp}}$ we write
\be
\om^{\rm{exp}}(t) \equiv \frac{\rho(t)}{\rc^{\rm{exp}}}
\ee
This definition implies that $\om^{\rm{exp}}=\om G_{\rm{exp}}/G(t)$ or
\be
\label{Gom}
G(t)=\frac{\om(t)}{\om^{\rm{exp}}(t)}\;G_{\rm{exp}}
\ee

In \cite{br2} we investigated the consequences of the conjecture that, for $k \to
0$, the RG trajectory for the dimensionless Newton constant $g(k) \equiv k^2
G(k)$ and cosmological constant $\lambda(k) \equiv \Lambda(k)/k^2$ runs into 
an infrared attractive fixed point $(g_{\ast}, \lambda_{\ast})$ with $g_\ast > 0$
and $\lambda_\ast > 0$. This means that the 
corresponding dimensionful quantities behave as $G(k)=g_{\ast}/k^2$, $\Lambda(k) = 
\lambda_{\ast} k^2$ for $k \to 0$. In the fixed point regime, the coupled 
system (\ref{eq:system})
for $K=0$ and with $k(t)=\xi/t$ was found to have an essentially unique 
attractor solution which governs the cosmology for $t \to \infty$:
\begin{subequations}
\label{solution}
\ba
\label{eq:scalefact}
&&a(t)=\left[\left(\frac{3}{8}\right)^2(1+w)^4 g_{\ast} \lambda_{\ast} \mathcal{M} 
\right]^{1/(3+3w)}t^{4/(3+3w)}\\
&&\label{eq:density} \rho(t)=\frac{8}{9 \pi (1+w)^4 g_{\ast}\lambda_{\ast}} 
\frac{1}{t^4}\\
&&G(t)=\frac{3}{8}(1+w)^2 g_{\ast} \lambda_{\ast} t^2\\
&&\label{eq:cosmconst} \Lambda(t)=\frac{8}{3 (1+w)^2} \frac{1}{t^2}
\ea
\end{subequations}
Here the conserved quantity
\be
\label{eq:M}
{\cal{M}} \equiv 8 \pi \rho(t) [a(t)]^{3+3w}=\textrm{const}
\ee
is the only free constant of integration. The system (\ref{eq:system}) also fixes the 
parameter $\xi$:
\be
\label{eq:csi}
\xi^2=\frac{8}{3(1+w)^2\lambda_{\ast}}
\ee
The cosmology (\ref{solution}) yields $\rho=\rho_{\Lambda}=\rc/2$, $\rho_{\rm tot}
=\rc$, or
\be
\om= \Omega_{\Lambda}=\frac{1}{2}, \quad \Omega_{\rm tot}=1
\ee
It is one of the most attractive consequences of the fixed point hypothesis 
that it leads unambiguously to these values of the relative densities. They 
are intriguingly close to, but not identical with,
the values favored by the analyses of the recent observations within standard
cosmology. We shall come back to this point later.

The Hubble parameter of the cosmology (\ref{solution}) is
\be
\label{hubble}
H(t) = \frac{4}{3+3w} \frac{1}{t},
\ee
and the deceleration parameter reads
\be
q \equiv -\frac{a \ddot a}{\dot a^2} = \frac{3w-1}{4}
\ee
We can use (\ref{hubble}) in order to compute the age of the Universe, $t_0$, in terms of 
the present Hubble constant $H(t_0) \equiv H_0$:
\be
t_0 = \frac{4 H_0^{-1}}{3+3w}  
\ee
Clearly, for the late universe, the most plausible equation of state is $p=0$, 
i.e. $w=0$. In this case the fixed point solution describes an accelerated 
expansion $a \propto t^{4/3}$ with the deceleration parameter $q=-1/4$ \cite{berto}. For the 
age of the Universe we obtain $t_0=\frac{4}{3}H_0^{-1}$. This is precisely 
twice the age one would obtain in standard cosmology with $\Lambda=0$ for 
the same value of $H_0$.

\subsection{Measuring $G(t_0)$ and $g_{\ast}\lambda_{\ast}$}
In the sequel we assume that the present Universe is described by the fixed 
point solution for $K=0$, eqs. (\ref{solution}), and that this fixed point 
behavior started at a certain transition time $t_{\rm tr}<t_0$. Later on we 
shall determine $t_{\rm tr}$ by postulating a plausible form of the RG 
trajectory running into the fixed point.

In this section we discuss how, at least in principle, one can determine the 
present cosmological Newton constant $G(t_0)$ and the product $g_{\ast} \lambda_{\ast}$ 
which characterizes the fixed point. Note that the attractor solution 
(\ref{solution}) depends only on this product but not on $g_{\ast}$ and $\lambda_{\ast}$ 
separately. 
The experience with the {\it ultraviolet} fixed point of QEG \cite{ol,ol2} suggests
that the product $g_{\ast} \lambda_{\ast}$ should be scheme 
independent (universal) while the factors $g_{\ast}$ and $\lambda_{\ast}$ are not.

Since the fixed point solution satisfies $\rho(t)=\rt/2$, we can use 
(\ref{eq:costante}) in order to express $G(t_0)$ in terms of observable 
quantities:
\be
\label{eq:G}
G(t_0)=\frac{3}{16 \pi}\; \frac{H_0^2}{\rho(t_0)}
\ee
By a very precise measurement of the Hubble constant and the matter energy 
density we can, in principle, determine $G(t_0)$ and compare it to the 
laboratory value $G_{\rm{exp}}$. It is clear, however, that before we can 
check whether $G(t_0) \ne G_{\rm{exp}}$, the experimental situation must 
improve considerably. (Note also that the RHS of (\ref{eq:G}) differs only by 
a factor of 2 from what one obtains in standard cosmology with $\Lambda=0$, 
$K=0$; in this case the prefactor is $3/8\pi$.)

Because $\om(t)=1/2$ holds throughout the fixed point regime, we can interpret 
a deviation of $G(t_0)$ from $G_{\rm {exp}}$ also in terms of a 
$\om^{\rm{exp}}(t_0)$-value which differs from $1/2$. From (\ref{Gom}) we 
obtain
\be
\label{eq:Gomexp}
G(t_0)=\frac{G_{\rm{exp}}}{2 \; \om^{\rm{exp}}(t_0)}
\ee

In order to get a first idea about the ratio $G(t_0)/G_{\rm{exp}}$ we consider the 
hypothetical extreme case in which there exists no dark matter at all.
Within the standard theory, the density of known forms of matter yields roughly
\be
\om^{\rm{exp}}(t_0) \gtrsim \frac{1}{100}
\ee
Interpreting this figure within our model we find that
$G(t_0)/G_{\rm{exp}} \lesssim 50$. Note that the ratio 
$G(t_0)/G_{\rm{exp}}$ cannot be many orders of magnitude larger than unity precisely 
because the present density is so close to the critical one.

As for the determination of $g_{\ast}\lambda_{\ast}$, we start from the relation
\be
g_{\ast}\lambda_{\ast}=G(k)\Lambda(k)=G(t)\Lambda(t)=\textrm{const}
\ee
which, in the fixed point regime, holds for any cutoff identification $k=k(t)$.
Thus, $g_{\ast}\lambda_{\ast}=G(t_0)\Lambda(t_0)$ where we may substitute
\be
\Lambda(t_0)=\frac{3}{2} \; H_0^2
\ee
which follows from (\ref{eq:lambda}) with $\Omega_{\Lambda}(t_0)=1/2$.
Therefore we can express $g_{\ast}\lambda_{\ast}$ in terms of observable quantities 
either as
\be
\label{eq:first}
g_{\ast}\lambda_{\ast}=\frac{3}{2}\; G(t_0)\; H_0^2
\ee
or as
\be
g_{\ast}\lambda_{\ast}=\frac{9}{32\pi}\; \frac{H_0^4}{\rho(t_0)}
\ee
We see that we can determine $g_{\ast}\lambda_{\ast}$ by measuring $H_0$ and $G(t_0)$ or
$\rho(t_0)$, respectively. Let us introduce the Hubble length $\ell_{\rm H}(t) 
\equiv 1/H(t)$ and the Planck length $\ell_{\rm Pl} = \sqrt {G_{\rm{exp}}}$, 
defined in the usual way in terms of the laboratory value of Newton's constant.
 In terms of these length scales (\ref{eq:first}) can be rewritten as
\be
g_{\ast}\lambda_{\ast}=\frac{3}{2}\; \frac{G(t_0)}
{G_{\rm{exp}}} \left( \frac{\ell_{\rm Pl}}{\ell_{\rm H}(t_0)}
\right)^2
\ee
Since  $\ell_{\rm Pl}/\ell_{\rm H}
(t_0)=O(10^{-60})$ 
the product $g_{\ast}\lambda_{\ast}$ is of the order $\left[ G(t_0)/G_{\rm exp}\right] \, 10^{-120}$. 
Most probably $G(t_0)$ and $G_{\rm exp}$ do not differ by many orders of
magnitude so that, roughly, $g_{\ast}\lambda_{\ast}=O(10^{-120})$. 

We emphasize that in our scenario the smallness of this number does not pose 
any finetuning problem as it does in standard cosmology. In fact, $g_{\ast}\lambda_{\ast}$ is a 
fixed and universally defined number which in principle can be computed from 
the RG equation. However, apart from being a difficult technical problem, this 
computation is possible only if we know the complete system of matter fields
in the Universe. The number $10^{-120}$ reflects specific properties of this 
matter system coupled to gravity rather than an initial condition.

\subsection{A natural extension of the model}\label{sec:exten}
Up to this point, our entire discussion was based upon the single hypothesis 
that, for $k \to 0$, the dimensionless couplings $g(k)$ and 
$\lambda(k)$ are attracted towards a fixed point $(g_{\ast}, 
\lambda_{\ast})$ at $g_{\ast}>0$ and $\lambda_{\ast}>0$. By virtue of the system 
(\ref{eq:system}) this information is sufficient in order to determine the 
cosmological evolution for $t \to \infty$.
In the $K=0$ - sector it was found to be given by the attractor solution 
(\ref{solution}). 
With this restricted knowledge about the RG trajectory, it is clearly 
impossible to determine how the Universe evolved towards the attractor, 
or to answer the question when the fixed point behavior set in.

In order to have a model which applies also before the transition to the fixed
point era we shall now generalize our hypothesis about the RG trajectory in a 
very ``minimal'', in fact the simplest possible way. 
We shall assume that the fixed point era of the cosmological evolution is 
preceded by an era during which $G$ and $\Lambda$ are approximately constant.

To be precise, our first hypothesis is that the relevant RG trajectory, at 
least at a qualitative level, can be approximated by\footnote{Eqs. (\ref{eq:Gk}),
(\ref{eq:Lambdak}) could, for instance, be regarded as a model for the final part 
of a possible crossover from the UV to the IR fixed point, similar to what happens
 in 2D gravity \cite{liouv}.}
\be
\label{eq:Gk}
G(k) = \left\{ \bay{r}
g_{\ast}/k^2 \qquad \textrm{for} \; k<k_{\rm tr}\\
G_< \qquad \textrm{for} \; k>k_{\rm tr} \eay \right.
\ee
\be
\label{eq:Lambdak}
\Lambda(k) = \left\{ \bay{r}
\lambda_{\ast}k^2 \qquad \textrm{for} \; k<k_{\rm tr}\\
\Lambda_< \qquad \textrm{for} \; k>k_{\rm tr} \eay \right.
\ee
We fix the constant values $G_<$ and $\Lambda_<$ in terms of the ``transition 
scale'' $k_{\rm tr}$ according to
\ba
\label{eq:Gm}
&&G_<=\frac{g_\ast}{k_{\rm tr}^2}\\
\label{eq:Lambdam}
&&\Lambda_<=\lambda_{\ast}k_{\rm tr}^2
\ea
This guarantees that $G(k)$ and $\Lambda(k)$ are continuous (but not 
differentiable) at the transition point. Clearly a more realistic RG trajectory
would be differentiable, and the transition from the regime with $G, \Lambda =
const$ to the fixed point regime would occur smoothly during a finite 
interval $\Delta k_{\rm tr}$ centered about $k_{\rm tr}$. The trajectory 
(\ref{eq:Gk}), (\ref{eq:Lambdak}) should always be understood as an 
idealization of a smooth RG trajectory interpolating between $G, \Lambda = 
const$ and the fixed point running.

Our second hypothesis concerns the identification of the cutoff in terms of 
dynamical variables which in general could be of the form $k=k(t, a(t), 
\dot a(t), \cdot \cdot \cdot)$. Guided by the discussion in \cite{br2} we assume that 
the identification $k=\xi/t$, $\xi>0$, is valid for all $k<k_{\rm tr}$, where 
$t_{\rm tr}=\xi/k_{\rm tr}$ denotes the time at which the transition occurs. (As we 
shall see, the scale $k_{\rm tr}$ or the transition time $t_{\rm tr}$ are not input 
parameters but rather are predicted by the model itself.)

It is easy to solve the coupled system (\ref{eq:system}) for $t>t_{\rm tr}$ and 
$t<t_{\rm tr}$, respectively. 
For $t>t_{\rm tr}$, the most general solution is the one parameter 
family of attractor solutions (\ref{solution}), with $\mathcal M$ being the 
only free parameter. For $t<t_{\rm tr}$, the unique solution (with $a(0)=0$) is
 the standard spatially flat FRW cosmology with a cosmological constant 
$\Lambda_<>0$. Its scale factor reads \cite{br1}
\be
\label{eq:scaleFRW}
a(t)=\left[ \frac{\mathcal{M}G_<}{2\Lambda_<}\left\{\textrm{cosh}\left[ (1+w)
\sqrt{3\Lambda_<} \;\; t\right]-1\right\} \right]^{1/(3+3w)}
\ee
and the matter density is
\be
\rho(t)=\frac{\mathcal{M}}{8 \pi [a(t)]^{3+3w}}
\ee
Note that the parameter $\mathcal{M}$ has the same value in both branches of the solution 
because it is related to a constant of motion of the system (\ref{eq:system}) by 
eq. (\ref{eq:M}).

As we discussed in detail in \cite{br1}, the very existence of solutions to the 
improved evolution equations (\ref{eq:system}) is a nontrivial issue because 
for a given RG 
trajectory and cutoff identification this system is overdetermined. While in 
the case at hand it is very easy to solve the equations for $t<t_{\rm tr}$ and 
$t>t_{\rm tr}$ separately, it is by no means guaranteed that, at $t=t_{\rm tr}$, 
the 
two solutions match in a physically acceptable way. Basically we are trying to 
adjust {\it one} parameter, $t_{\rm tr}$, in such a way that {\it four} 
functions,
$a(t), \rho(t), G(t), \Lambda(t)$ become continuous at $t_{\rm tr}$.

It is 
important here that in reality the transition from the classical FRW regime 
to the fixed point regime does not take place instantaneously, but during a 
finite interval of time, $\Delta t_{\rm tr}$, which is centered about $t_{\rm 
tr}$. 
This transition period reflects the fact that the true RG trajectory is 
differentiable, and that there is a smooth transition from $G,\Lambda= 
const$ to $G \propto 1/k^2$, $\Lambda \propto k^2$ during a finite but, 
by assumption, small interval $\Delta k_{\rm tr}$. The classical FRW solution is valid only for $t \lesssim  
t_{\rm tr} - \Delta t_{\rm tr}/2$, while the fixed point solution applies for 
$t \gtrsim t_{\rm tr} + \Delta t_{\rm tr}/2$. During the transition regime
\be
t_{\rm tr} - \Delta t_{\rm tr}/2 \; \lesssim \; t \; \lesssim \; t_{\rm tr} + 
\Delta t_{\rm tr}/2
\ee
the evolution is much more complicated, and we are not going to describe it 
explicitly. We assume that the system (\ref{eq:system}) applies also in the 
transition period,
and that a more realistic version of the RG trajectory (\ref{eq:Gk}), 
(\ref{eq:Lambdak}) gives rise to a 
continuous and differentiable solution for all four functions. The 
interpolating solution is likely to exist because during the transition the 
cutoff identification may well be much more complicated than $k \propto 1/t$, 
the actual function $k=k(t, a(t), \dot a(t), \cdot \cdot \cdot)$ being fixed to 
some extent by the requirement that (\ref{eq:system}) is consistent.

Thus our model provides an idealized description which ignores the transition 
regime, $\Delta t_{\rm tr} \to 0$. This is a sensible approximation if 
$\Delta t_{\rm tr}$ is much smaller than $t_{\rm tr}$. Since in reality the 
functions 
$a, \rho, G$ and $\Lambda$ do have a certain variation between $t_{\rm tr} - 
\Delta t_{\rm tr}/2$ and $t_{\rm tr} + \Delta t_{\rm tr}/2$, it is clear, then,
that in the model these functions should be allowed to have some moderate discontinuity at 
$t=t_{\rm tr}$. More precisely, we would consider it reasonable if 
$a(t\nearrow 
t_{\rm tr})/a(t\searrow t_{\rm tr})$, say, is a small number close to unity. 
On the other hand, if this ratio should turn out many orders of magnitude smaller 
than unity, say, then this means that the Universe undergoes an enormous 
``inflation'' during the transition regime. In this case a crucial qualitative 
element would be missing from our description if we just consider the classical
FRW solution and the fixed point solution explicitly. Hence the idealized 
model would not be very useful probably.

Let us now look at the properties of the extended IRFP model in detail. To start with, we 
assume that (\ref{eq:Gm}) and (\ref{eq:Lambdam}) hold true exactly so that 
$G(t)$ and $\Lambda(t)$ are continuous at the matching point. 
Eqs. (\ref{eq:Gm}) and (\ref{eq:Lambdam}) imply that
\be
G_< \Lambda_<=g_{\ast} \lambda_{\ast}
\ee
This means that even if one allows for small discontinuities of $G$ and 
$\Lambda$, the product $G_< \Lambda_<$ is of the order of $10^{-120}$. As the 
cosmological constant problem in its original form (``Why is $\Lambda\ll
m_{\rm Pl}^2$?'') is related to the smallness of this number, we see that the 
mechanism which has achieved this tiny value must have been operative already 
{\it before} the FRW era with $G_<$ and $\Lambda_<$ has started. (Later on we 
shall assume that the FRW era began before the time of nucleosynthesis.)

Using $t_{\rm tr}=\xi/k_{\rm tr}$ with $\xi$ given by (\ref{eq:csi}) we obtain 
for the time of the transition:
\be
t_{\rm tr}^2= \frac{\xi^2}{k_{\rm tr}^2}=\frac{8}{3 (1+w)^2 \lambda_{\ast} k_{\rm tr}^2}=
\frac{8}{3 (1+w)^2 \Lambda_<}
\ee
Hence, in terms of $\Lambda_<$ or $G_<$,
\ba
\label{eq:trtime}
t_{\rm tr}&=&\frac{1}{(1+w)} \; \sqrt{\frac{8}{3 \Lambda_<}}\\ \nonumber
&=&\frac{1}{(1+w)} \; \sqrt{\frac{8 G_<}{3 g_{\ast} \lambda_{\ast}}}
\ea
Another useful representation of $t_{\rm tr}$ in terms of the age of the 
Universe 
can be obtained by eliminating $g_{\ast} \lambda_{\ast}$ from the second line of 
(\ref{eq:trtime}) by virtue of (\ref{eq:first}) with $H_0=4/[3(1+w)t_0]$. The 
result is surprisingly simple:
\be
\label{eq:trtime2}
t_{\rm tr}= t_0 \sqrt{\frac{G_<}{G(t_0)}} 
\ee

Now we are in a position to calculate $a(t_{\rm tr})$ both from the fixed point
solution $a(t) \equiv a_{\rm{FP}}(t)$ of (\ref{eq:scalefact}) and the 
classical FRW solution $a(t) \equiv a_{\rm FRW}(t)$ in (\ref{eq:scaleFRW}). 
With (\ref{eq:trtime}) we obtain
\be
a_{\rm{FP}}(t_{\rm tr})= \left[ \frac {{\mathcal M}G_<}{\Lambda_<} \right]^{1/(3+3w)}
\ee
and
\be
a_{\rm FRW}(t_{\rm tr})= C_w \left[ \frac {{\mathcal M}G_<}{\Lambda_<} 
\right]^{1/(3+3w)}=C_w \; a_{\rm{FP}}(t_{\rm tr})
\ee
where
\be
C_w \equiv \left( \frac {\rm{cosh}(\sqrt 8)-1}{2} \right)^ {1/(3+3w)}
\ee

In the following we continue the discussion for the most relevant equation of state 
$p=0$, i.e. $w=0$.

In this case the ratio of the two scale factors, $C_w$, has the numerical value
$C_0 \approx 1.55$. It is encouraging to see that this number is indeed 
relatively close to unity. As $C_0 > 1$ the FRW scale factor at $t=t_{\rm tr}$ 
is {\it larger} than the one of the fixed point regime. We discussed above 
that, strictly speaking, the FRW solution $a_{\rm FRW}(t)$ is valid only for $t 
\lesssim t_{\rm tr}-\Delta t_{\rm tr}/2$ 
and the fixed point solution $a_{\rm{FP}}(t)$ applies 
only for $t\gtrsim t_{\rm tr}+\Delta t_{\rm tr}/2$. During the transition a more 
complicated solution interpolates smoothly between $a_{\rm FRW}$ and $a_{\rm{FP}}$. 
The simplest possible behavior of the interpolating solution one could 
imagine is that $a(t)$ stays approximately constant between $t_{\rm tr}-
\Delta t_{\rm tr}/2$ and $t_{\rm tr}+\Delta t_{\rm tr}/2$. If one then chooses $\Delta 
t_{\rm tr}$ such that
\be
\label{eq:contin}
a_{\rm FRW}(t_{\rm tr}-\Delta t_{\rm tr}/2)=a_{\rm{FP}}(t_{\rm tr}+\Delta t_{\rm tr}/2)
\ee
the resulting scale factor is continuous and nondecreasing. Clearly, also a 
more complicated behavior of the exact solution is conceivable, but, even 
without knowing the details of the exact solution, the condition 
(\ref{eq:contin}) gives us a first idea about the size of $\Delta t_{\rm tr}$. 
Being interested in an approximate estimate only, it is actually more convenient to define $\Delta 
t_{\rm tr}$ in a slightly different way, by $a_{\rm FRW}(t_{\rm tr})=a_{\rm{FP}}(t_{\rm tr}
+\Delta t_{\rm tr})$, which boils down to
\be
C_0 = \frac{a_{\rm{FP}}(t_{\rm tr}+\Delta t_{\rm tr})}{a_{\rm{FP}}(t_{\rm tr})}= 
\left( \frac{t_{\rm tr}+\Delta t_{\rm tr}}{t_{\rm tr}} \right)^{4/3}
\ee
whence
\be
\label{eq:ratio}
\frac{\Delta t_{\rm tr}}{t_{\rm tr}}=C_0^{3/4}-1 \approx 0.39
\ee
Even though $\Delta t_{\rm tr}$ is not very much smaller than $t_{\rm tr}$ 
itself, the ratio (\ref{eq:ratio}) indicates that there can very well be a 
window in which the idealized model is useful for a first orientation.

Note that, as a consequence of the exact relation (\ref{eq:M}), the 
discontinuity of $\rho(t)$ is necessarily of the same order of magnitude as 
that of $a(t)$.

From now on we shall be slightly more specific and assume that the FRW era with
$G=G_<$ and $\Lambda=\Lambda_<$ has started before the time of primordial 
nucleosynthesis already (with $w=1/3$ then). This assumption allows us to fix 
the value of $G_<$. In fact, from the successful predictions of nucleosynthesis
 theory we know that the value of the cosmological Newton's constant during this 
epoch was very close to the laboratory value: $G(t_{\rm nucl})=
G_{\rm{exp}}$. Since, by assumption, $G(t_{\rm nucl})=G_<$, this entails
\be
G_<=G_{\rm{exp}}
\ee
With this identification, the transition time (\ref{eq:trtime2}) reads
\be
\label{eq:trtime3}
t_{\rm tr}= t_0 \; \sqrt{\frac{G_{\rm{exp}}}{G(t_0)}} 
\ee
This equation is quite remarkable. It relates the time when the fixed point 
behavior started to the ratio of the laboratory value of Newton's constant 
and the cosmological Newton's constant. Using (\ref{eq:Gomexp}) we can reexpress 
this ratio in terms of $\om^{\rm{exp}}$:
\be
\label{eq:trtime4}
t_{\rm tr}= t_0 \; \sqrt{2 \, \om^{\rm{exp}}(t_0)}
\ee
Eqs. (\ref{eq:trtime3}) and (\ref{eq:trtime4}) are our main results
of this section.

It is convenient to characterize $t_{\rm tr}$ by the corresponding redshift 
$z_{\rm tr}$. We have
\be
\label{eq:zeta}
1+z_{\rm tr}= \frac{a(t_0)}{a(t_{\rm tr})}=\left(\frac{t_0}{t_{\rm tr}}\right)^{4/3}
\ee
and
\be
1+z_{\rm tr}= \left(\frac{G(t_0)}{G_{\rm{exp}}}\right)^{2/3}=
\left[ 2 \, \om^{\rm{exp}}(t_0) \right]^{-2/3}
\ee

Which values of $z_{\rm tr}$ could we expect? The assumption $\om^{\rm{exp}}(t_0) 
\gtrsim 1/100$ yields $t_{\rm tr}\gtrsim t_0/7$ and $z_{\rm tr}\lesssim 12$
, for instance. A distinguished period during the ``recent''
evolution of the Universe is the era of galaxy formation at about $z\approx 3$, say.
If we speculate that the fixed point behavior started during this era
we have
\be
\label{eq:bound}
z_{\rm tr} \approx 3, \qquad t_{\rm tr}\approx \frac{t_0}{2.8}
\ee
This scenario corresponds to $\om^{\rm{exp}}(t_0)\gtrsim 1/16$.
It has been shown in ref. \cite{pert} that there are no growing 
``large'' wavelength (which probably means super Hubble size)
density perturbations in the fixed point regime. However, ``small''-scale
density perturbations can grow even in the fixed point regime. Hence 
$z_{\rm tr}>3$ is not necessarily excluded by the theory of structure formation.

We expect that if $z_{\rm tr}$ is too small and the transition happened only 
recently, the fixed point solution cannot provide an accurate description of 
the present Universe because the transient phenomena of the transition region 
are still visible to some extent. Let us write $\Delta t_{\rm{FP}} \equiv t_0 - 
t_{\rm tr}$ for the time which, according to the model, elapsed since the onset
 of the fixed point behavior. As a rough estimate we can say that the present 
Universe is already well within the fixed point regime, and the transient 
effects have become insignificant, provided $\Delta t_{\rm{FP}} \gtrsim 
\Delta t_{\rm tr}$. We rewrite this inequality as $\Delta t_{\rm{FP}}/t_{\rm tr} \gtrsim 
\Delta t_{\rm tr}/t_{\rm tr}$. Then its LHS is given by (\ref{eq:zeta}),
\be
\frac{\Delta t_{\rm{FP}}}{t_{\rm tr}}=\left(1+z_{\rm tr}\right)^{3/4}-1,
\ee
and (\ref{eq:ratio}) yields $C_0^{3/4}-1$ for the RHS. Hence we are already well within the 
fixed point regime if $1+z_{\rm tr} \gtrsim C_0$ or
\be
\label{eq:bound2}
z_{\rm tr} \gtrsim 0.55
\ee
Thus we have a wide range of $z_{\rm tr}$-values where 
the model can be applied consistently.

The estimate (\ref{eq:bound2}) is of a rather intriguing order of magnitude. In fact,
the most distant supernova analyzed in \cite{perl} and \cite{riess} has a redshift of about $z \approx 1$,
while the highest redshift bin for the radio sources in \cite{gurv} (see Section \ref{sec:radio}) is centered about 
$z=3.6$. 
Thus, if the transition occurred relatively late, around $z=0.5$, say, the 
current data might already tell us something about the transition 
from the FRW to the fixed point regime. This question will be discussed in detail
in Sections \ref{sec:snia} and \ref{sec:radio}.
\subsection{Summary of the extended IRFP model}
We investigated a simple cosmological model in which the IR fixed point regime 
of \cite{br2} is preceded by a classical FRW era with constant values of $G$
and $\Lambda$. While the hypothesis of this FRW era does not introduce any 
unknown new parameter, it allows for a determination of the time when the 
Universe entered the fixed point regime. In a nutshell, this time, 
$t_{\rm tr}$, obtains from the intersection point of the parabola 
$G_{\rm{FP}} \propto 
t^2$ with the straight line $G_{\rm FRW}(t)=G_<$. The transition time $t_{\rm tr}$ 
turns 
out to be given by the ratio of the cosmological and laboratory value of 
Newton's constant or, equivalently, by the matter energy density $\om^
{\rm{exp}}(t_0)$. If this density is of the order of the critical density, 
the ratio $t_0/t_{\rm tr}$ is close to unity, i.e. the transition happened only
 ``recently''. In this sense the model solves the ``coincidence problem'', the question why 
$\rho$ and $\rho_{\Lambda}$ happen to be of the same order of magnitude 
precisely today. In our model we have $\rho/\rho_{\Lambda}=1$ at any time later
than  $t_{\rm tr}$.
\section{Type Ia supernovae and the IRFP cosmology}
\label{sec:snia}
Recent supernova data from several sources 
(Supernova Cosmology Project \cite{perl},
High Redshift Supernova Team \cite{riess}) provide a valuable tool in 
cosmological hypothesis testing, since they contain information about the luminosity distance 
 function $d_L(z)$. In this section we shall apply the standard 
$\chi^2$ test in order to analyze the
viability of the IRFP hypothesis.

In the Appendix, an alternative approach based on median statistics \cite{gott,avel} and 
a Bayesian model selection criterion is discussed in detail.

\subsection{The data set}\label{sec:data}
The data set used in all of our analyses consists of the 18 low redshift type Ia SNe from
the Calan-Tololo survey \cite{hamuy} plus 74 high redshift type Ia SNe, 
namely:
\begin{itemize}
\item 42 SNe discovered by the SCP, whose measured redshifts and apparent 
magnitudes are published in Perlmutter et al. \cite{perl};
\item 32 SNe discovered by the HzST, who reported their measured
 redshifts and distance moduli in Riess et al. \cite{riess}.
\end{itemize}
This gives a total of 92 supernovae. The oldest, most distant supernova is SN 1997ck, with 
a redshift of $0.97 \equiv z_{\rm max}$.
\subsection{Luminosity distance-redshift relations}\label{sec:lumz}
Let us assume that, at time $t_1$, a distant galaxy emits light which is 
detected on Earth at $t_0$. The luminosity distance $d_L$ to this galaxy is 
defined by the relation
\be
{\mathcal F} = \frac{{\mathcal L}}{4 \pi d_L^2}
\ee
where $\mathcal{L}$ is the absolute luminosity of the source and $\mathcal{F}$
is the measured flux. From the Robertson-Walker kinematics one obtains
\be
\label{eq:lumdist}
d_L= \frac{a^2(t_0)}{a(t_1)}\; S \left( \int_{t_1}^{t_0} \frac{dt}{a(t)} 
\right)
\ee
where
\be
S(x) = \left\{\begin{array}{ll}
\sin x & \quad \textrm{for} \; K=+1\\
x & \quad \textrm{for} \; K=\phantom{+}0\\
\sinh x & \quad \textrm{for} \; K=-1 \end{array} \right.
\ee
Using $a(t_0)/a(t_1)=1+z$ we may rewrite the integral in (\ref{eq:lumdist}) in 
terms of the redshift $z$:
\be
\int_{t_1}^{t_0} dt \frac{a_0}{a(t)} = \int_0^z dz' \frac {1}{H(z')}
\ee
(The subscript ``$0$'' on any quantity denotes its value at $t=t_0$.) In 
particular for $K=0$,
\be
\label{eq:hubble}
d_L(z)=\frac{1+z}{H_0} \int_0^z dz' \frac {H_0}{H(z')}
\ee
In order to calculate $H_0/H(z')$, a further distinction is necessary:
\begin{enumerate}
\item The {\it IRFP cosmology} has $K=0$ and is given by 
(\ref{solution}) where we shall set $w=0$ from now on. The corresponding 
luminosity distance as a function of the redshift of the source reads
\be
\label{eq:fp}
d_L(z)= \frac{4(1+z)}{H_0} \Big [(1+z)^{1/4}-1 \Big ]
\ee
Eq. (\ref{eq:fp}) is obtained either by inserting $a \propto t^{4/3}$ into 
(\ref{eq:lumdist}) and eliminating $t_1$ in favor of $z$, or by using 
(\ref{eq:hubble}) with
\be
\label{eq:hubblefp}
H(z)=H_0 \, (1+z)^{3/4}
\ee
which follows from (\ref{hubble}).
\item In standard {\it FRW cosmology} one has the familiar relationship
\be
\label{eq:hubblefrw}
H(z)=H_0 \Big [(1+z)^3 \Omega_{\rm M0}+\Omega_{\Lambda 0}+(1+z)^2 \Omega_{\rm K0} \Big ]^{1/2}
\ee
which can be used to derive
\be
\label{eq:fp2}
d_L(z)=\frac{(1+z)}{H_0 \sqrt{|\Omega_{K0}|}} S \left(\sqrt{|\Omega_{K0}|} 
\int_0^z dz' \Big [(1+z')^3 \Omega_{\rm M0}+\Omega_{\Lambda 0}+(1+z')^2
 \Omega_{K0} \Big ]^{-1/2} \right)
\ee
As for the actual form of the function $S$, 
eq. (\ref{eq:kol}) shows that the cases $K=+1$, $0$ and $-1$ are realized if 
$\Omega_{\rm K0} < 0$, $\Omega_{\rm K0}=0$ and $\Omega_{\rm K0} > 0$, respectively.
\item In the {\it extended IRFP model} introduced in  
subsection \ref{sec:exten} we combine a $K=0$ FRW cosmology, valid for $t<t_{\rm tr}$ 
($z>z_{\rm tr}$), 
with the fixed point cosmology which applies for $t>t_{\rm tr}$ 
($z<z_{\rm tr}$). For 
arguments $z<z_{\rm tr}$, the corresponding function $d_L(z)$ is given by (\ref{eq:fp}). For 
arguments $z>z_{\rm tr}$ we divide the $z'$-integral of (\ref{eq:hubble}) in a 
IRFP plus a FRW piece:
\be\label{eq:fp3}
d_L(z)=\frac{1+z}{H_0} \left\{ \int_0^{z_{\rm tr}} dz' \left[\frac {H_0}{H(z')}\right]
_{\rm IRFP}+ \int_{z_{\rm tr}}^z dz' \left[\frac {H_0}{H(z')}\right]_{\rm FRW} \right\}
\ee
The integrand of the IRFP piece is given by (\ref{eq:hubblefp}) and for the FRW 
term we use (\ref{eq:hubblefrw}) with $\Omega_{\rm K0}=0$:
\be
\label{eq:extend}
d_L(z)=\frac{1+z}{H_0} \left\{ \int_0^{z_{\rm tr}} dz' (1+z')^{-3/4}+ \int_{z_{\rm tr}}^z 
dz' \Big [(1+z')^3 \widetilde{\Omega}_{\rm M0} + \widetilde{\Omega}_{\Lambda 0}\Big ]^ {-1/2}
\right \}
\ee
The quantities $\widetilde{\Omega}_{\rm M0}$ and $\widetilde{\Omega}_{\Lambda 0}=1
-\widetilde{\Omega}_{\rm M0}$ are not the physical densities at $t=t_0$, these are 
$\Omega_{\rm M0}=\Omega_{\Lambda 0}=1/2$, but rather fictitious values which would
 result from using the standard FRW differential equations also for $t$ between
 $t_{\rm tr}$ and $t_0$. The quantities $\widetilde{\Omega}_{\rm M0}$ and 
$\widetilde{\Omega}_{\Lambda 0}$ are determined by the requirement that the 
FRW- and the IRFP-solution match at $z=z_{\rm tr}$; they depend on $z_{\rm tr}$ 
therefore.

In the present context it is most convenient to define $t_{\rm tr}$ by the 
requirement that $\om(t)$ and $\oa(t)$ are continuous at the 
transition.\footnote{The $t_{\rm tr}$ thus defined coincides with the one from subsection
\ref{sec:exten} within $\Delta t_{\rm tr}$.} Then the specific FRW theory which connects to the fixed point 
cosmology is singled out by the requirement
\be
\Omega_{\Lambda}^{\rm FRW}(t_{\rm tr})=\Omega_{\Lambda}^{\rm IRFP}(t_{\rm tr})=1/2
\ee
and similarly for $\om$. On the FRW side we can use the general result \footnote{Eq. 
(\ref{eq:olambda}) follows from (\ref{eq:hubblefrw}) by noting that $\rl=const$ in the 
standard case, and that $\rc \propto H^2$.}
\be
\label{eq:olambda}
\Omega_{\Lambda}(z)=\Omega_{\Lambda 0} \Big [(1+z)^3 \Omega_{\rm M0}+\Omega_{\Lambda 0}+
(1+z)^2 \Omega_{\rm K0} \Big ]^{-1}
\ee
in order to relate $\Omega_{\Lambda}^{\rm FRW}(t_{\rm tr})$ to $\widetilde
{\Omega}_{\Lambda 0}$. Letting $z=z_{\rm tr}$, $\Omega_{\Lambda}(z_{\rm tr})=1/2$, 
$\Omega_{\Lambda0} \to \widetilde{\Omega}_{\Lambda 0}$, $\Omega_{\rm M0} \to 
\widetilde{\Omega}_{\rm M0}$, eq. (\ref{eq:olambda}) yields
\be
\widetilde{\Omega}_{\Lambda 0}=(1+z_{\rm tr})^3 \, \widetilde{\Omega}_{\rm M0}
\ee
which, with $\widetilde{\Omega}_{\rm M0}=1-\widetilde{\Omega}_{\Lambda 0}$, 
leads to
\ba\label{eq:final1}
\widetilde{\Omega}_{\rm M 0}=\frac{1}{1+(1+z_{\rm tr})^3}\\
\label{eq:final2}
\widetilde{\Omega}_{\Lambda 0}=\frac{(1+z_{\rm tr})^3}{1+(1+z_{\rm tr})^3}
\ea
Using (\ref{eq:final1}) and (\ref{eq:final2}) in (\ref{eq:extend}) we obtain 
the final result for the 
luminosity distance in the case $z>z_{\rm tr}$:
\ba\label{eq:lumi}
d_L(z)&=&\frac{1+z}{H_0}  \Big\{ 4(1+z_{\rm tr})^{1/4}-4 \nonumber\\
&& +  \Big [1+(1+z_{\rm tr})^3 \Big ]^{1/2} 
\int_{z_{\rm tr}}^z dz' \Big [(1+z')^3+(1+z_{\rm tr})^3 \Big ]^{-1/2} \Big\}
\ea
\end{enumerate}

It is customary to express luminosity distances in terms of the distance 
modulus $\mu_0 \equiv m(z)-M$ according to
\be\label{eq:mag}
m(z)=M+5 \log_{10} \left ( \frac{d_L(z)}{1 \;{\rm Mpc}} \right) +25
\ee
where $m$ and $M$ are the apparent and absolute magnitude, respectively. 
Measuring $d_L$ in units of the present Hubble radius, $1/H_0$, the dependence 
of the apparent magnitude on the ``Hubble-free'' luminosity distance $H_0 d_L$ 
reads
\be \label{eq:mag2}
m(z)={ \mathcal M} +5 \log_{10} \Big ( H_0 \, d_L(z) \Big)
\ee
where\footnote{There should arise no confusion with the invariant (\ref{eq:M}) with the same 
name $\mathcal M$ which is also customary.}
\be
{\mathcal M} \equiv M - 5 \log_{10} \left ( H_0 \cdot 1 \; {\rm Mpc} \right) + 25
\ee
is constant within an ensemble of standard candles.

The HzST results are presented in terms of the distance modulus, while 
the SCP published the estimated effective {\it B}-band magnitude
$m_{\rm B}^{\rm{eff}}$ which relates to the HzST results through
\begin{equation}
\label{eq:mag3}
m_{\rm B}^{\rm{eff}} = M_{\rm B} + \mu _0\,
\end{equation}
where $M_{\rm B}$ is the peak {\it B}-band absolute magnitude of a standard 
type Ia supernova.
We chose to translate the HzST data from distance moduli into apparent 
magnitudes, leaving the SCP data unchanged.

In the expression for the apparent magnitude, $M_{\rm B}$ and $H_0$ appear only in 
the combination ${\mathcal{M}}_{\rm B}=M_{\rm B}-5 \log_{10} (H_0 \cdot 1 \rm Mpc) +25$.  
We have used the value of ${\mathcal{M}}_{\rm B}$ that is obtained
from the analysis of the low-redshift SN sample of the Calan-Tololo survey \cite{hamuy} alone.
If $z \ll 1$, the expression (\ref{eq:mag2}) for the apparent magnitude 
becomes
\be\label{eq:mag4}
m_{\rm B}(z)={\mathcal{M}}_{\rm B}+5 \log_{10} z 
\ee
Fitting the measured apparent magnitudes to this formula yields 
${\mathcal{M}}_{\rm B}=-3.32$. 

The total data set now consists of 92 measured apparent magnitudes $m_{\rm B}^{\rm meas}$, 
henceforth denoted $m_i^{\rm meas}$, $i=1,\cdots,92$, which must be compared 
to the theoretical expectation for the 
Hubble-free expression for the apparent magnitude,
\be
\label{eq:appmag}
m_{\rm B}(z)={\mathcal{M}}_{\rm B}+5 \log _{10} \Big (H_0 \, d_L(z) \Big) 
\ee
\begin{figure}[!h]
\psfig{figure=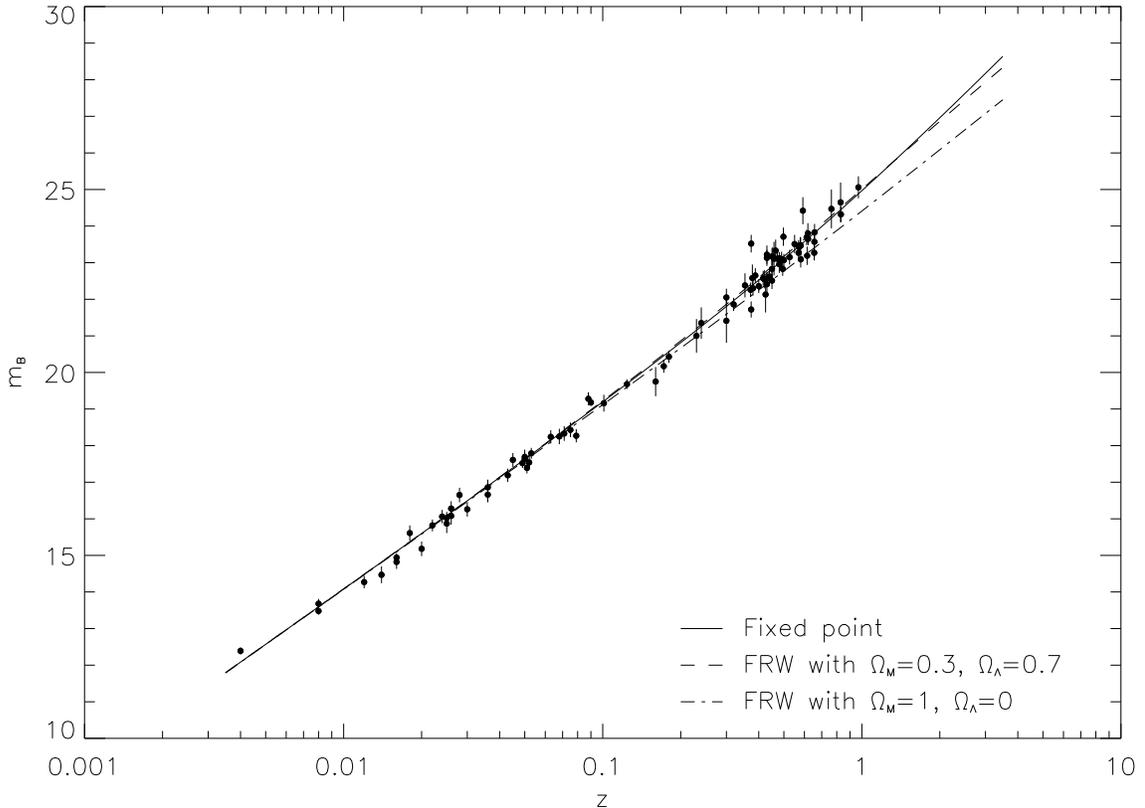,width=1.0\textwidth}
\caption{The measured apparent magnitudes of the supernovae as a function of their redshift.
The continuous line represents the prediction of the IRFP cosmology, the dashed one is 
the best-fit FRW model, 
and the dot-dashed line is a flat FRW model with zero cosmological constant.
\label{fig1}}
\end{figure}

Fig. \ref{fig1} displays the apparent magnitudes $m_i^{\rm meas}$ as a function of the 
corresponding measured redshifts $z_i$. The error bars on the data points indicate the statistical
errors $\sigma_i$ of $m_i^{\rm meas}$ estimated by SCP and HzST, respectively. Fig. \ref{fig1} 
also shows the theoretical curve (\ref{eq:appmag}) with the function $d_L(z)$ pertaining to 
three different cosmological models: the IRFP model (without a FRW part), a FRW model with 
$\Omega_{\rm M0}=0.3$, $\Omega_{\Lambda 0}=0.7$, and a FRW model with $\Omega_{\rm M0}=1$, 
$\Omega_{\Lambda 0}=0$. The second model is the 
one which, in the framework of standard FRW cosmologies, fits the data best. The question 
is whether the IRFP leads to an even better fit.

\subsection{$\chi^2$ analysis of the currently available data}\label{par:results}

The standard $\chi^2$ test is the most common choice in order to quantify the agreement 
between the predictions of some cosmological model and the supernova data. From the triples 
$(z_i, m_i^{\rm meas}, \sigma_i)$, $i=1,\cdots,92$, provided by the dataset described in 
\ref{sec:data} we have to compute

\be
\label{eq:chi2}
\chi^2 = \sum_{i=1}^{92} \left[\frac{m_i^{\rm meas}-m^{\rm theor}_i}{\sigma_i}\right]^2
\ee
where $m^{\rm theor}_i \equiv m_{\rm B}(z_i)$ is the theoretical value predicted by eq. (\ref{eq:appmag}) 
with the function $d_L(z)$ computed within a specific model. As for the fixed point model, we 
must distinguish the cases $z_{\rm tr} > z_{\rm max}$ and $z_{\rm tr} < z_{\rm max}$. In the 
former, the Universe had already entered the fixed point regime by the time the oldest supernova exploded; 
in the latter, some of the supernova events took place during the preceding FRW era, and their photons 
detected by our telescopes have witnessed the transition to the fixed point regime.

If $z_{\rm tr} > z_{\rm max}$ then $d_L(z_i)$ is given by (\ref{eq:fp}) for all supernovae. The evaluation 
of the sum (\ref{eq:chi2}) yields a value of $\chi^2$ per degree of freedom ($\chi^2/\rm dof$) of $1.59$. In 
Table \ref{tab:table0} we compare this number to the corresponding $\chi^2$ values for three competing FRW 
cosmologies:

\begin{itemize}
\item The best-fit FRW model with vanishing cosmological constant and $\Omega_{\rm M0}$ anywhere in the 
interval $[0,3]$;
\item The best-fit FRW model with $\Omega_{\rm M0} \in [0,3]$ and $\Omega_{\Lambda 0} \in [-1,3]$ such that 
$\Omega_{\rm M0} + \Omega_{\Lambda 0} = 1$, i.e. spacetime is spatially flat;
\item The best-fit FRW model with $\Omega_{\rm M0} \in [0,3]$ and $\Omega_{\Lambda 0} \in [-1,3]$ arbitrary.
\end{itemize}

We see that the $\chi^2/$dof value of the IRFP cosmology is about the same as those
for the FRW models. Taking the numbers in Table \ref{tab:table0} at face value, the 
fixed point model performs even slightly better than its FRW competitors. This is 
a very encouraging result, in particular since, contrary to the FRW models, 
{\it the fixed point model has no free parameters} which could be adjusted.

\begin{table}[!hb]
\bc
\caption{The value of $\chi^2$ per degree of freedom for several cosmological models (the errors on the best-fit parameter values correspond to the 1-$\sigma$ confidence region).
\label{tab:table0}}
\begin{ruledtabular}
\begin{tabular}{lcc}
\bf Model \footnote{The variation intervals for $\omz$ and $\oaz$ in the fits are [0,3] and [-1,3], respectively. The transition redshift $z_{\rm tr}$ goes from $0$ to $0.97$, the redshift of the farthest SN in the set.}                                                       &\bf $\chi^2/{\rm dof}$   & \bf Parameter values 	\\
								 &			   & \bf at minimum		\\
\hline
Fixed point model ($z_{\rm tr}>z_{\rm max}$)                       	&       1.59       &		-		\\
Extended fixed point model						&	1.60	   &	$z_{\rm tr}=0.17_{-0.10}^{+0.80}$		\\
Best-fit FRW with $\oaz=0$              &       1.62       &	$\omz=0.00^{+0.05}$			\\
Best-fit FRW with $\oaz+\omz=1$		&   	1.60       &    $\omz=0.40^{+0.10}_{-0.05}$			\\
Best-fit FRW		         	&       1.62       &    $\omz=0.25^{+0.65}_{-0.25}$, $\oaz=0.40^{+0.75}_{-0.40}$	\\
\end{tabular}
\end{ruledtabular}
\ec
\end{table}

Up to here we assumed that $z_{\rm tr} > z_{\rm max}$ so that even the oldest 
supernova in the dataset would not "know" about the onset of the fixed point 
epoch. Assuming the opposite case $z_{\rm tr} < z_{\rm max}$ now, some of the 
older supernovae exploded already during the FRW era which, according to the 
extended IRFP model, preceded the fixed point epoch. The function $d_L(z)$ 
of the extended IRFP model depends parametrically on $z_{\rm tr}$. For 
supernovae with $z_i < z_{\rm tr}$ the value of $d_L(z_i)$ is given by equation 
(\ref{eq:fp}), while for $z_i > z_{\rm tr}$ eq. (\ref{eq:lumi}) must be used. As a 
consequence, $\chi^2$ is a function of $z_{\rm tr}$ now, and in principle one 
could hope to extract the value of $z_{\rm tr}$ from the data by looking where 
$\chi^2=\chi^2(z_{\rm tr})$ is minimum. In the ideal case the function $\chi^2
(z_{\rm tr})$ would have a pronounced minimum at some value of $z_{\rm tr}$, 
and we would then be entitled to conclude that the fixed point regime started 
at this redshift most probably.

\begin{figure}[!t]
\psfig{figure=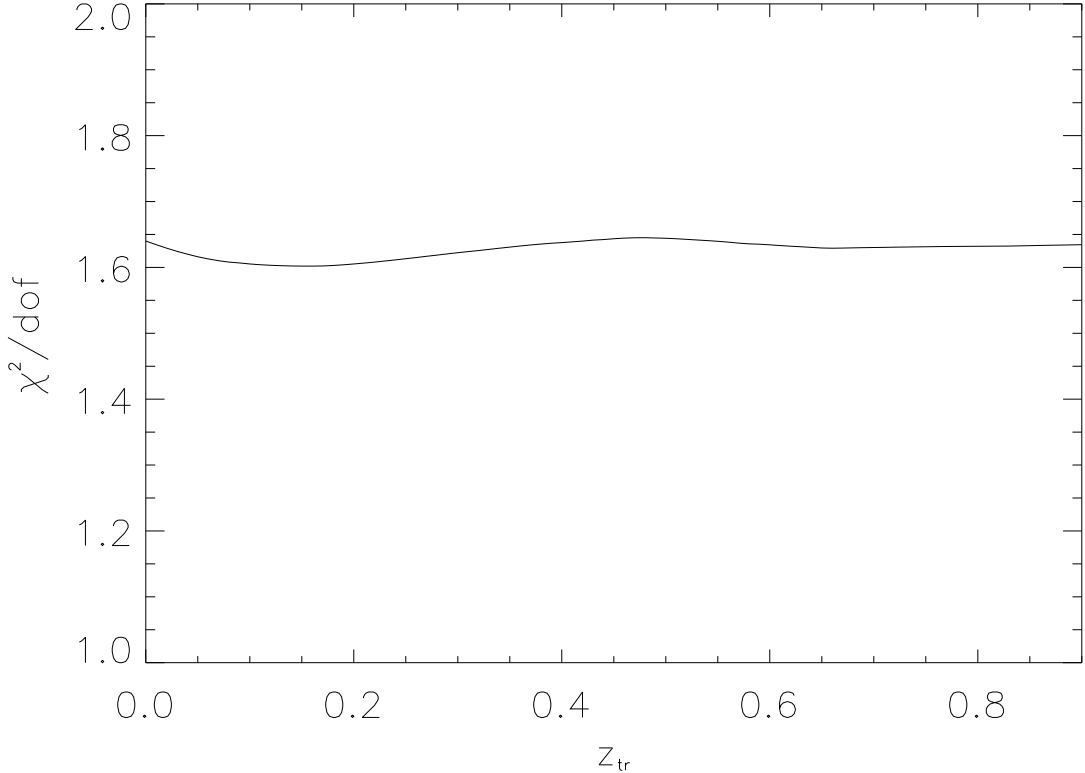,width=1.0\textwidth}
\caption{The $\chi^2$ value of the extended IRFP model as a function of 
the redshift $z_{\rm tr}$ obtained from the currently 
available data set. 
\label{fig2}}
\end{figure}

Using again the data set of subsection \ref{sec:data} we obtain the function 
$\chi^2=\chi^2(z_{\rm tr})$ which is plotted in Fig. \ref{fig2}. Obviously 
there is no strongly preferred value of $z_{\rm tr}$. There seems to be a 
minimum at $z_{\rm tr} \approx 0.18$, but it is too shallow to derive any 
statistically significant conclusion from it.

Thus we must conclude that the statistical quality of the currently available 
supernova data is not good enough in order to discriminate between the fixed 
point model proper (without a FRW era) and the one-parameter family, parametrized 
by $z_{\rm tr}$, of extended fixed point models.

\subsection{$\chi^2$ analysis of Monte Carlo data for SNAP}

It is an important question whether it would be possible to detect a "smoking gun" 
of the transition to the IRFP regime in the supernova data if a better statistics 
were available.
In fact, the {\it SNAP} ({\it Supernova Acceleration Probe}) satellite 
is aimed to improve this statistics at a significant level. Its goal is to obtain a super
data set more than one order of magnitude larger than the
currently available one, with an improved control
over systematic errors, to redshifts up to about $z\approx 1.7$.
We therefore decided to simulate the expected results in order to forecast the
impact of this enlarged data set on the constraints for $z_{\rm tr}$.
In particular we used the SNOC code, developed by A.Goobar {\it et al.}
\cite{goobar}, which takes into account gravitational interactions (lensing) and extinction
by dust, both in the host galaxy and along the line-of-sight, and modified it by implementing 
our new distance-redshift formula.

\begin{figure}[!t]
\psfig{figure=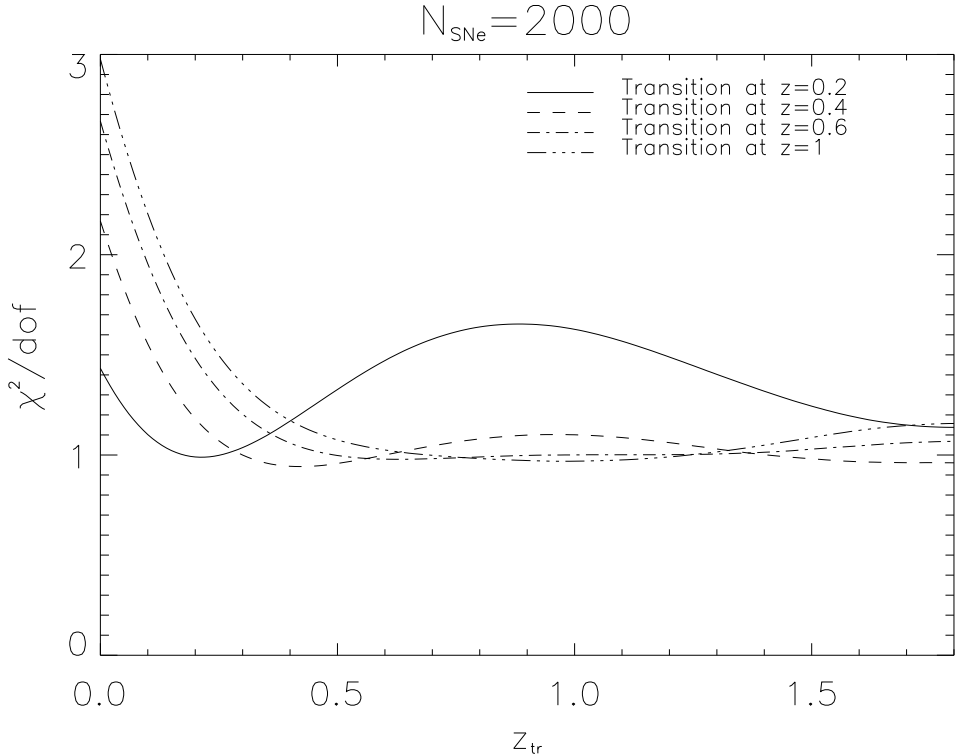,width= 15cm}
\caption{The function $\chi^2(z_{\rm tr})$ from a simulated data set
of 2000 supernovae for various values of the transition redshift  
$z_{\rm tr}^{\rm simul}$.
\label{fig2bis}}
\end{figure}

The modified SNOC code based upon the extended IRFP model was used to generate 4 
different Monte Carlo data sets, each consisting of 2000 type Ia supernovae in 
the redshift interval $[0.1,1.8]$. The 4 simulations differed with respect to the 
transition redshift $z_{\rm tr}^{\rm simul}$ of the underlying IRFP model; we 
chose $z_{\rm tr}^{\rm simul}=0.2,0.4,0.6$ and $1.0$, respectively. We then 
performed a $\chi^2$-analysis of the 4 data sets as we did with the "real" data 
in the previous subsection. Fig. \ref{fig2bis} shows the resulting $\chi^2$ as a function 
of $z_{\rm tr}$ (not to be confused with $z_{\rm tr}^{\rm simul}$). We find that, 
if the transition took place late enough, at a redshift of $0.2$, say, there is a 
clear minimum in the $\chi^2(z_{\rm tr})$ curve, and it should be possible to 
determine $z_{\rm tr}$ from the SNAP data\footnote{In view of the discussion 
which led to the estimate (\ref{eq:bound2}) it is clear, however, that a quantitatively correct 
description of this late transition would require a more sophisticated model than 
that of Section \ref{sec:model}.}. On the other hand, if it occurred much earlier, at a 
redshift of $1.0$, say, the $\chi^2(z_{\rm tr})$ curve does not prefer any 
specific value of $z_{\rm tr}$. For $z_{\rm tr} \gtrsim 0.5$ this curve is almost flat. 
This is related to the fact that, if $z_{\rm tr} \gtrsim 1$, the luminosity distance 
functions $d_L(z)$ of (\ref{eq:fp}) and (\ref{eq:lumi}) are virtually identical for arguments 
$z \in [0,2]$, say.

The lesson we learn from this Monte Carlo investigation is that, if we improve the 
statistics by a factor of about 10, there is a significant hope of either actually 
detecting the transition to the fixed point regime if $z_{\rm tr} \lesssim 0.5$, or at 
least of putting a lower bound on $z_{\rm tr}$ if $z_{\rm tr} \gtrsim 0.5$.

In the Appendix and in Section V we resume the analysis of the currently available ``real'' supernova 
data. Since they are anyhow not sufficient to determine $z_{\rm tr}$ we assume that 
$z_{\rm tr} > z_{\rm max}$ there.

\section{Compact radio sources and the IRFP cosmology}
\label{sec:radio}
In the previous section we have shown how a class of astronomical objects of well-known 
intrinsic properties can help probing the cosmic evolution via the luminosity distance-redshift 
relation.

A similar, yet independent test is based on the angular diameter distance-redshift relation. 
The {\it angular diameter distance} $d_A(z)$ of an object with redshift $z$ and proper diameter $l$ 
which subtends an angle $\theta$ as seen by a terrestrial observer is defined as
\be
\label{eq:def}
d_A(z) \equiv \frac{l}{\theta} = \frac{D/H_0}{\theta}
\ee
where we have introduced the characteristic angular size $D \equiv l H_0$ which is to be interpreted
as an angle given in milliarcseconds (mas). 
By measuring the object's redshift and $\theta$-angle one can compute its value of $d_A(z)$
and compare it to some particular cosmological model, provided its diameter $l$ is known. 
By comparing the predictions of several different 
models, the one which best fits the data can then be estabilished.
Of course, a class of {\it standard rods} (objects with the same characteristic extension $l$)
is needed in order for this determination of $d_A(z)$ to be viable. 
Recently \cite{gurv}, with some manipulation and redshift-binning, a set of 330 compact radio sources has been applied to 
this purpose, thus giving the opportunity to constrain the free parameters of several 
cosmologies, from FRW and power-law models to cardassian expansion scenarios \cite{gurv,jain,zhu}.

This section intends to address precisely the issue of confronting the IRFP 
model with the new data. The statistical framework is analogous to the one adopted in \ref{sec:snia}: 
both the proper and extended versions of the IRFP model are compared to FRW cosmologies via the 
$\chi^2$ test. The $\chi^2$ test for the extended IRFP model is applied to the data set in order to give an 
estimate of the transition time $t_{\rm tr}$.

\subsection{The data set}
The class of standard objects to be used as distance indicators is compact 
($\lesssim 100 \; \rm pc$) radio sources. Specifically, the set consists of 330 sources with redshifts $z_i$ in the interval 
$[0.011,4.72]$ and angular diameters $\theta_i$, extracted 
from 5 GHz VLBI contour maps in the literature, as reported by Gurvits et al. 
\cite{gurv}.

Several reasons point out why these are likely to be standard objects: first, evolutionary effects are expected to be 
negligible with respect to galaxies or extended two-lobes radio sources, since the characteristic time 
scale of the former 
(some ten years) is very tiny compared to the age of the universe, or to the Hubble time $H_0^{-1}$. Second, while the 
physics of extended systems is more likely to be influenced by the different properties 
of the intergalactic medium encountered at 
different redshifts, the features of compact sources can in principle be described by a few parameters, like the mass of the 
central black hole, the magnitude of the magnetic field and of the angular momentum, and a few more.

Three tasks have thus to be completed before the data set can be applied to the cosmological analyses:
\begin{itemize}
\item[(i)] To control and put limits on the few basic parameters governing the radio source, 
so as to have a collection of 
objects which is as uniform as possible;
\item[(ii)] To homogenize the set, which has been assembled from contour maps published by 
different collaborations, rather
 than from a unique set of observations;
\item[(iii)] To investigate possible evolutionary effects (linear size - redshift dependence), as well as 
linear size - luminosity and linear size - spectral index dependences, and minimize their influence 
on the cosmological implications.
\end{itemize}

To these aims, Gurvits at al. \cite{gurv} restrict the set by retaining only those sources with 
luminosity $Lh^2 \geq 10^{26} \; \rm W/Hz$ 
and spectral index $-0.38 \leq \alpha \leq 0.18$. Since a moderate dependence of $\alpha$ on 
the angular size seems to be 
estabilished (see Fig. 7 in \cite{gurv}), the last relation has the further 
effect of constraining the dimensions of the sources.

The new group of 145 sources is then binned into 12 redshift intervals, and a median redshift 
and angular size is calculated 
for each one, together with the standard deviation $\sigma_i$. This procedure has the additional effect 
of reducing the influence of particularly large objects in the set, and 
providing simple means to deal with unresolved sources (see cases J, L and S in \cite{gurv} for details).

\begin{figure}
\psfig{figure=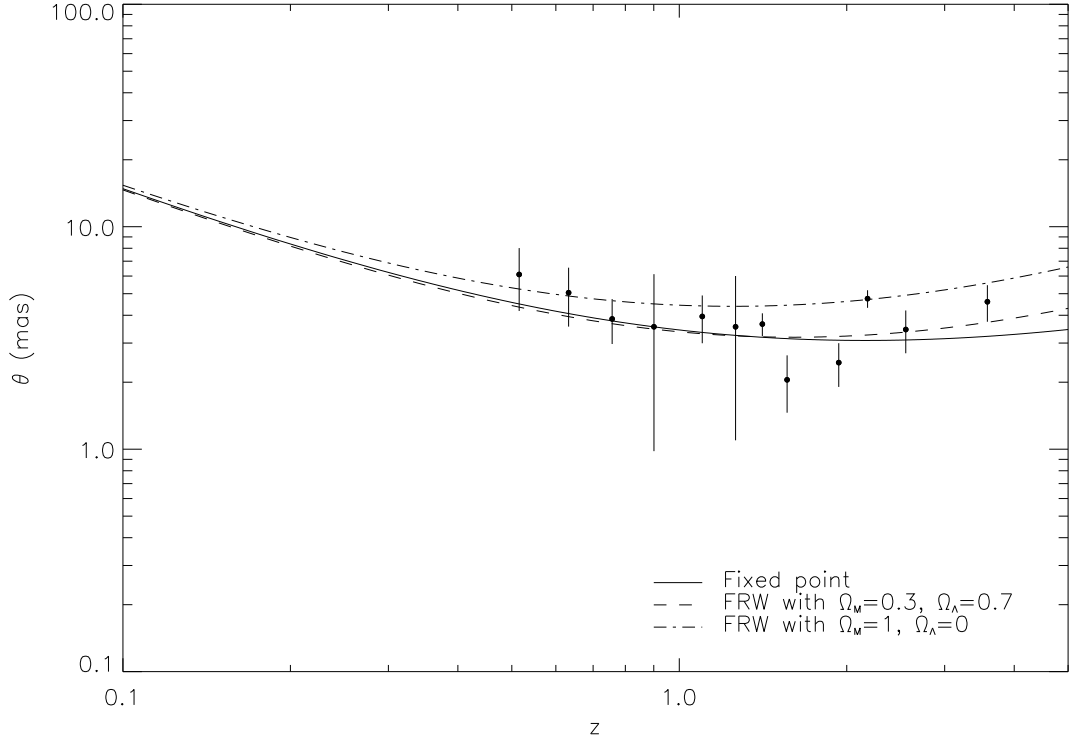,width= 15cm}
\caption{The 12 data points (with corresponding uncertainties) and the theoretical predictions 
of the proper IRFP model and of the FRW models with $\om=0.3, \oa=0.7$ and with $\om=1, \oa=0$. 
The angular size $D$ is assumed to equal its best-fit value $D = 1.30$ .
\label{fig:punti}}
\end{figure}

The final data set consists of 12 triples $(z_i,\theta_i,\sigma_i)$ in the redshift interval $[0.52,3.6]$. 
The data points, 
together with the three curves representing the proper IRFP prediction and the 
FRW models with $\om=0.3, \oa=0.7$ and with $\om=1, \oa=0$, are presented in Fig. \ref{fig:punti}.

\subsection{The angular diameter distance-redshift relation}
From (\ref{eq:def}), following simple reasoning, one can easily show that in any spatially flat
FRW spacetime
\be\label{inlu}
d_A(z)=\frac{d_L(z)}{(1+z)^2}= \frac{1}{H_0(1+z)} \int_0^z dz'\frac{H_0}{H(z')}
\ee
In \ref{sec:lumz} we demonstrated that, in the extended IRFP model, for 
$z$ greater than $z_{\rm tr}$,
\ba
\int_0^z dz' \frac{H_0}{H(z')}  = 4(1+z_{\rm tr})^{1/4}-4
+  \Big [1+(1+z_{\rm tr})^3 \Big ]^{1/2} 
\int_{z_{\rm tr}}^z dz' \Big [(1+z')^3+(1+z_{\rm tr})^3 \Big ]^{-1/2} 
\ea
while, for $z < z_{\rm tr}$,
\be
\int_0^z dz' \frac{H_0}{H(z')} = 4 \Big[ (1+z)^{1/4}-1 \Big]
\ee
It is evident from the above formulas that $d_A(z)$ also depends on the transition time, 
as long as the transition does not 
occur at a redshift $z_{\rm tr}$ which is greater than the maximum $z$ of the sources 
in the data set ($z_{\rm max} \approx 3.6$, 
see below.). We can therefore investigate the possibility of a transition to the fixed point 
regime in the redshift interval 
$[0,3.6]$. This window is much larger than the one allowed by the supernova set adopted in 
\ref{sec:snia}.

\subsection{$\chi^2$ analysis and determination of the transition time}
\begin{figure}
\psfig{figure=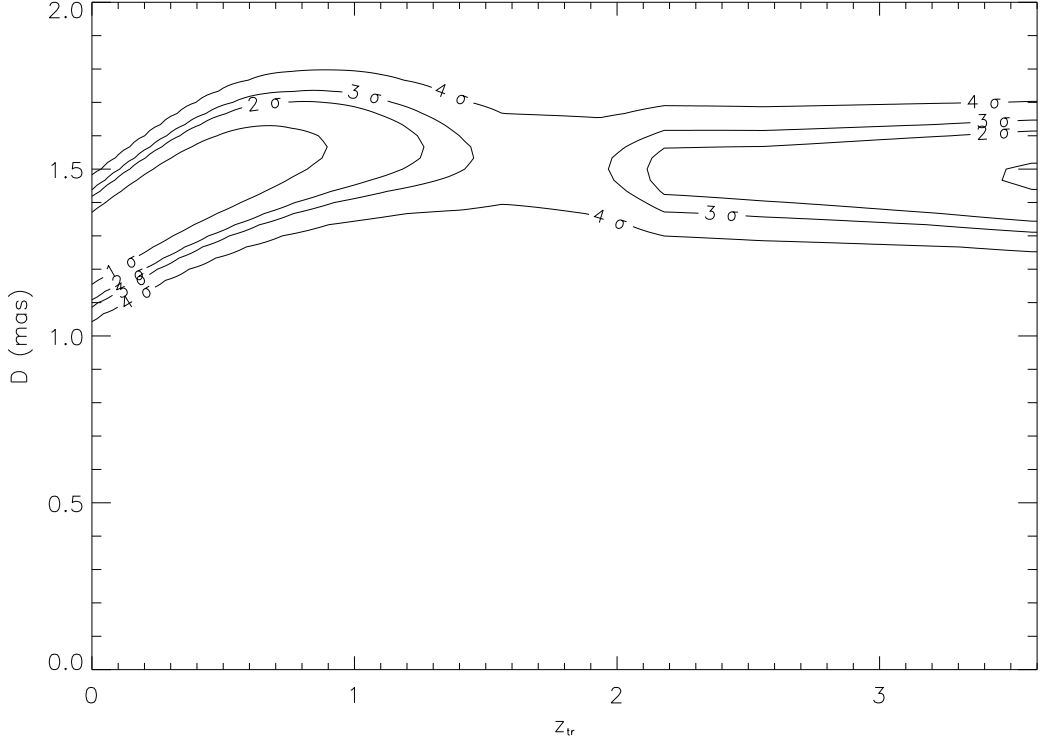,width= 15cm}
\caption{Contour levels for the probability $P(z_{\rm tr},D)$. 
\label{fig:1}}
\end{figure}

\begin{figure}
\psfig{figure=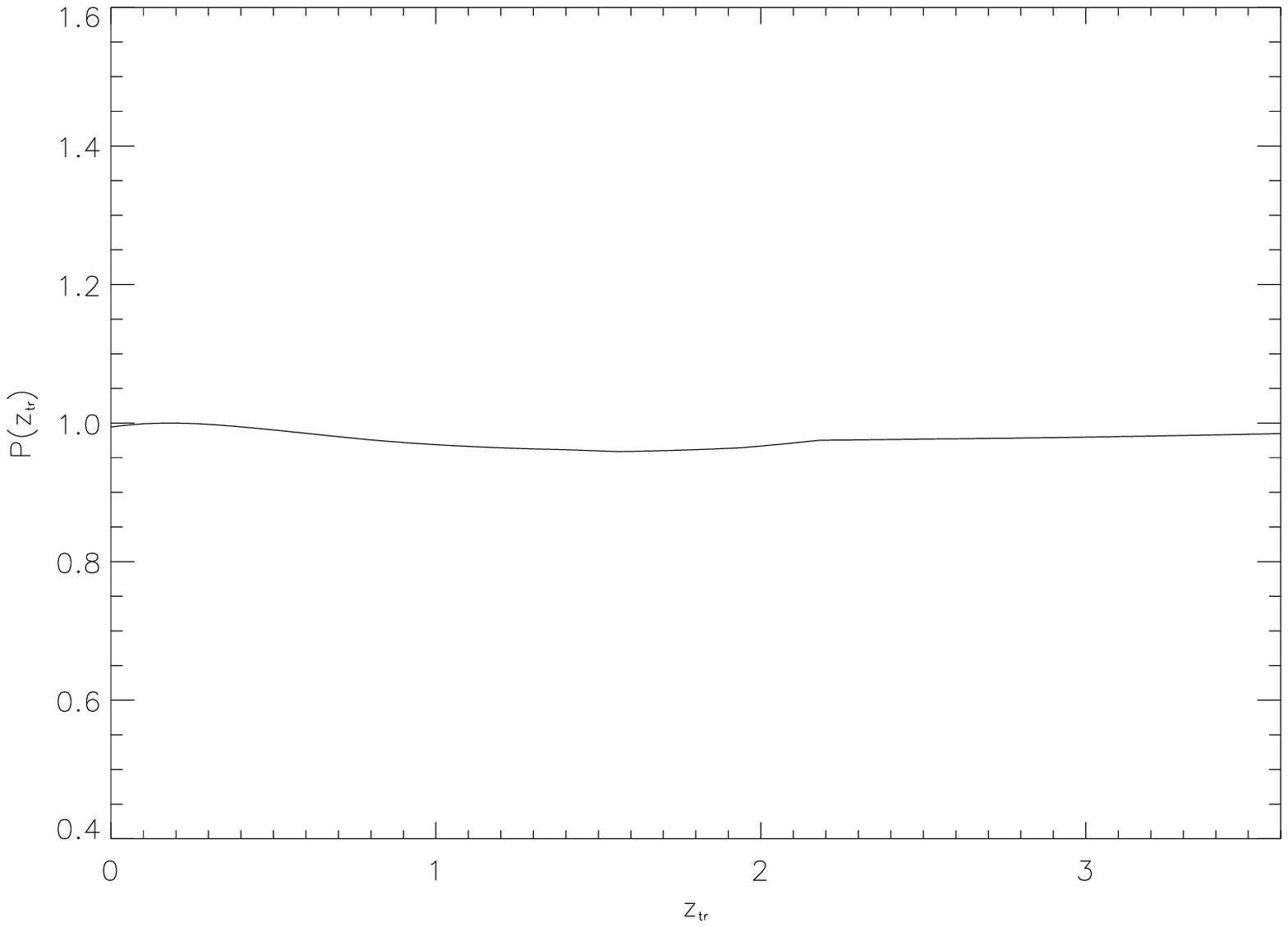,width= 15cm}
\caption{The marginalized probability $\widetilde{P}(z_{\rm tr})$. A shallow maximum 
can be seen at $z_{\rm tr}=0.16$.
\label{fig:2}}
\end{figure}

A $\chi^2$ function can thus be constructed in the usual manner:
\be
\label{eq:chi}
\chi^2(z_{\rm tr},D)= \sum_{i=1}^{12} \left [ \frac{\theta(z_i;z_{\rm tr})-\theta_i}{\sigma_i} \right ] ^2
\ee
Here $\theta_i$ stands for the observed values of the angular size with errors $\sigma_i$, and 
$$ \theta(z_i;z_{\rm tr}) = \frac{l}{d_A(z_i)}=\frac{D}{H_0 \; d_A(z_i)} $$ 
is the theoretical value.
Let us note that the $\chi^2$ value is also a function of the ``Hubble free'' diameter $D$ of the source 
(it does not depend on $H_0$ though; see Eq.(\ref{inlu}) for $d_A(z)$). 
The neatest way to solve this problem is to calculate a 2-dimensional 
$\chi^2$ in the parameter space of $z_{\rm tr}$ and $D$, and then marginalize over $D$ to make 
the results independent of its value \cite{jain}.
The level contours for the probability $P(z_{\rm tr},D)$, associated to $\chi^2(z_{\rm tr},D)$ are 
shown in Fig. \ref{fig:1}. $P(z_{\rm tr},D)$ is maximum at $(z_{\rm tr}=0.08, D=1.3 \; \rm mas)$, 
where $\chi^2$ has a 
corresponding value of 1.86.
In Table \ref{tab:radio}, this result is compared to other cosmologies.
Marginalizing over $D$, i.e. defining a new probability as
\be
\widetilde{P}(z_{\rm tr})=\frac{\int_0^2 dD \; 
P(z_{\rm tr},D) }{{\rm max}\left [ P(z_{\rm tr},D) \right ]}
\ee
one obtains the probability distribution for $z_{\rm tr}$ alone, as shown in Fig. \ref{fig:2}.

Clearly, the situation is even more problematic than for SNe Ia: the probability $\widetilde{P}(z_{\rm tr})$ 
hardly seems to vary, and a determination of $z_{\rm tr}$ is by no means possible with the current data.
However, again we may conclude that the IRFP model proper performs as well as standard cosmology.
\begin{table}[!hb]
\bc
\caption{The value of $\chi^2$ per degree of freedom for several cosmological models (the errors on the best-fit parameter values correspond to the 1-$\sigma$ confidence region).
\label{tab:radio}}
\begin{ruledtabular}
\begin{tabular}{lcc}
\bf Model \footnote{The variation intervals for $\omz$ and $\oaz$ in the fits are [0,3] and [-1,3], respectively. The transition redshift $z_{\rm tr}$ goes from $0$ to $4$.}                                                       &\bf $\chi^2/{\rm dof}$   & \bf Parameter values 	\\
								 &			   & \bf at minimum		\\
\hline
Fixed point model ($z_{\rm tr}>z_{\rm max}$)                       	&       1.88       &		$D=1.47_{-0.07}^{+0.06}$		\\
Extended fixed point model						&	1.86	   &	$z_{\rm tr}=0.08_{-0.08}^{+0.80}, D=1.30_{-0.13}^{+0.30}$		\\
Best-fit FRW with $\oaz=0$              &       1.91       &	$\omz=0.55_{0.20}^{+0.20}, D=1.20_{-0.30}^{+0.30}$			\\
Best-fit FRW with $\oaz+\omz=1$		&   	1.86       &    $\omz=0.60^{+0.15}_{-0.15}, D=1.20_{-0.30}^{+0.30}$			\\
Best-fit FRW		         	&       1.91       &    $\omz=0.15^{+1.65}_{-0.05}$, $\Omega_\Lambda=1.20^{+0.15}_{-2.20}$, 
$D=1.80^{+0.30}_{-0.90}$	\\
\end{tabular}
\end{ruledtabular}
\ec
\end{table}

\section{Beyond the IRFP model: general power law solutions}
In this section we are going to put the fixed point cosmology into a more 
general perspective by ``embedding'' it into a broader class of cosmologies 
with a time dependent $G$ and $\Lambda$. Let us give up for a moment the 
idea that the $t$-dependence of $G(t)$ and $\Lambda(t)$ stems from an 
underlying RG trajectory via some identification $k \equiv k(t)$. Then 
we are left with the system of equations 
(\ref{eq:IFE},\ref{eq:tensor},\ref{eq:BI}) 
without (\ref{eq:RGflow}). This system has been studied in 
the literature already long ago \cite{sys1}. The problem is that 
(\ref{eq:IFE},\ref{eq:tensor},\ref{eq:BI}) is underdetermined, so that in order to 
obtain a unique solution one has to ``invent'' some additional condition 
on $a, \rho, G$ and $\Lambda$ on an {\it ad hoc} basis. For instance, 
without providing a deeper explanation, it was assumed \cite{sys1} that Newton's 
constant follows a power law $G(t) \propto t^n$ with an arbitrary, not 
necessarily integer exponent $n$. With this additional equation, the 
system (\ref{eq:IFE},\ref{eq:tensor},\ref{eq:BI}) has the following 
solution for $K=0$:
\begin{subequations}
\label{eq:pl}
\ba
&&a(t) = \Big [\frac{3(1+w)^2}{2(n+2)}\;{\cal M}\; C\Big ]^{1/(3+3w)}\;t^{(n+2)/(3+3w)}\label{eq:4a}\\[2mm]
&&\rho(t) = \frac{(n+2)}{12\pi\;(1+w)^2\;C}\;\frac{1}{t^{n+2}}\label{eq:4b}\\[2mm]
&&G(t) = C\; t^n\label{eq:4c}\\[2mm]
&&\Lambda(t) = \frac{n(n+2)}{3(1+w)^2}\;\frac{1}{t^2}\label{eq:4d}
\ea
\end{subequations}
Actually (\ref{eq:pl}) describes a 2-parameter family of solutions labeled 
by the parameters $\mathcal M$ and $C$. The fixed point cosmology (\ref{solution}) 
is the special case of (\ref{eq:pl}) which is obtained by setting $n=2$ and 
$C=3(1+w)^2 g_\ast \lambda_\ast / 8$. In fact, the exponent $n=2$ is an unambiguous 
prediction of the fixed point hypothesis. It obtains not only for the simple cutoff 
identification $k \propto 1/t$ but even for an arbitrary function $k=k(t)$. This 
can be seen as follows. On the one hand we have, in the fixed point regime, 
$G(t) \Lambda(t)=G(k) \Lambda(k)=g_\ast \lambda_\ast$; on the other hand, eqs. 
(\ref{eq:4c}) and (\ref{eq:4d}) yield $G(t) \Lambda(t)=\rm const \ne 0$ if, and only if, $n=2$.

The luminosity distance for the models (\ref{eq:pl}) is easily worked out. It is 
independent of $\mathcal M$ and $C$, but it does depend on the exponent $n$. For a 
power law expansion $a(t) \propto t^\alpha$ we have in general
\be
\label{eq:pllum}
d_L(z)=\left ( \frac{\alpha}{\alpha-1} \right) \frac{(1+z)}{H_0} \left 
[ (1+z)^{1-1/\alpha}-1 \right ]
\ee
and (\ref{eq:4a}) yields in particular
\be
\label{eq:expon}
\alpha=\frac{n+2}{3+3w}
\ee

\begin{figure}[ht]
\psfig{figure=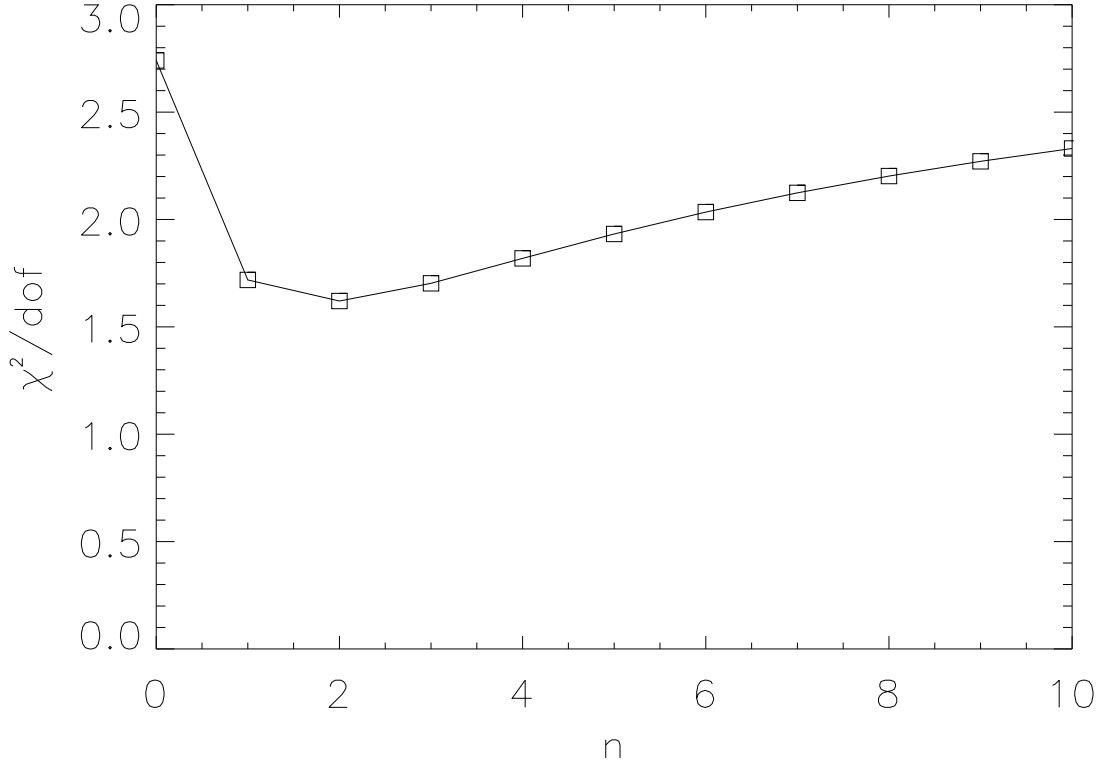,width=1.0\textwidth}
\caption{The $\chi^2$ values of the general power law models as a function of the
exponent $n$.
\label{fig:n}}
\end{figure}

We used the currently available supernova data in order to test the cosmologies 
(\ref{eq:pl}). Proceeding as in Section \ref{par:results}, we performed a $\chi^2$ 
analysis where $m_i^{\rm theor}$ was computed from the luminosity distance 
(\ref{eq:pllum}), (\ref{eq:expon}) with $w=0$, i.e. for $\alpha=(n+2)/3$. Fig. 
\ref{fig:n} displays the resulting $\chi^2$ values as a function of the exponent $n$.
Quite remarkably, we observe a clear minimum of $\chi^2$ at $n=2$, which is precisely 
the exponent predicted by the fixed point model. Stated differently, in the space of 
all cosmological models with a power law expansion $a \propto t^\alpha$ it seems to 
be exactly the $t^{4/3}$-expansion of the IRFP model which fits the supernova data best.
\section{Constraints on $z_{\rm tr}$ from the CMBR first acoustic peak}
The recent data from WMAP have determined
the position of the first acoustic peak in the CMBR spectrum with great accuracy \cite{wmap}. 
The spectrum depends on the complicated 
interplay between spacetime geometry and microphysics of the recombination era but
the position of the first Doppler peak is essentially a measure of the angular size of the sound horizon,
$\Delta \theta_s$, at the recombination epoch ($z_{\rm rec}\approx 1300$). 
One has  
\be\label{peak1}
\ell_1 \propto \frac{1}{\Delta \theta_s} {\Big |}_{z=z_{\rm rec}} = \frac{d_A (z_{\rm rec})}{d^S_H(z_{\rm rec})} 
\ee
where $d_A (z_{\rm rec})$ is the angular diameter distance to the surface of last scattering,
and $d^S_H(z_{\rm rec})$
is the proper radius of the sound horizon at decoupling \cite{weinp}. 
The precise proportionality factor on the LHS of (\ref{peak1})
is dermined by the microscopic model of the recombination.  
In standard FRW cosmology ${\Delta \theta_s}$ ranges from $\approx 0.85^\circ$ for a simple $\Lambda=0$ model, 
to about $\approx 0.60^\circ$, in a more refined calculation \cite{angle}. Thus, strictly speaking
the ${\Delta \theta_s}$-value inferred from a given measured multipole number $\ell_1$ is model 
dependent, but the model dependence changes ${\Delta \theta_s}$ at most by a factor of order 
unity. Taking it for granted that ${\Delta \theta_s}$ is about one degree or slightly less, { i.e.}
of the order of magnitude predicted by standard cosmology, we can derive an upper bound on 
$z_{\rm tr}$ in the following way. 

The key observation is that both $d_A(z_{\rm rec})$ and $d_H^S(z_{\rm rec})$ tend to {\it increase} when we
replace a piece of FRW evolution by the FP evolution. For $z_{\rm tr}$ small enough, their ratio 
${\Delta \theta_s}$ stays almost constant. 

As for $d_A(z_{\rm rec})$, Eq.(80) with (\ref{eq:pllum}) provides us with $d_A(z)$ for a power law
expansion $a(t)\propto t^\alpha$. We can use this formula for a comparison of the $t^{4/3}$-FP-evolution
and the $t^{2/3}$-FRW-evolution where we neglect the cosmological constant for a first orientation. 
In the extreme case of a transition during the recombination era, $z_{\rm tr}=z_{\rm rec}=1300$,
one finds
\be\label{factor}
d_A(z_{\rm rec})\Big |_{\alpha = 4/3} \approx 10.3 \; d_A(z_{\rm rec})\Big |_{\alpha = 2/3}
\ee
We conclude that for realistic redshifts $z_{\rm tr}\ll z_{\rm rec}$ the value of $d_A(z_{\rm rec})$
is bounded above by about 10 times its value in standard cosmology.

As for $d_H^S(z_{\rm rec})$, we recall that in any $(\Omega_{\rm M0},\Omega_{\rm \Lambda 0})$-FRW model
the sound horizon at redshift $z\gg \Omega_{\rm M0},\Omega_{\rm \Lambda 0}$ has proper radius
\be\label{89a}
d_H^S(z) \approx \frac{2 c_s^{\rm eff}}{H_0 \sqrt{\Omega_{\rm M0}}}\; (1+z)^{-3/2}
\ee
where $c_{\rm eff}$ is an effective speed of sound. In the extended FP model, the FRW cosmology
preceding the FP era is characterized by the densities $\widetilde{\Omega}_{\rm M0}$
and $\widetilde{\Omega}_{\rm \Lambda 0}$ of Eqs.(69), (70). This allows us to compare 
$d_H^S(z_{\rm rec})$ in the extended FP model to the corresponding value in the 
$\Omega_{\rm M0}=1$, $\alpha=2/3$ FRW model:
\be\label{factor2}
d^S_H(z_{\rm rec})\Big |_{\rm FP} = \sqrt{1+(1+z_{\rm tr})^3} \; \; d^S_H(z_{\rm rec})\Big |_{\alpha = 2/3}
\ee
Combining (\ref{factor}) to (\ref{factor2}) we observe that the increase of $d_A(z_{\rm rec})$ certainly
cannot compensate for the increase of $d_H^S(z_{\rm rec})$ so as to keep 
${\Delta \theta_s}$ unaltered if the square root in (\ref{factor2}) is larger than 10.3.
The condition $\sqrt{1+(1+z_{\rm tr})^3} < 10.3$
leads to un upper bound for the transition redshift:
\be
z_{\rm tr} < 3.7
\ee 
This is a remarkable order of magnitude. It indicates that the transition should have taken place
at or after the era of structure formation. This conclusion is in accord with the picture arising from
the analysis of density perturbations \cite{pert}. The actual value of $z_{\rm tr}$ is much 
smaller probably. For a more precise estimate the details of the microphysics should be taken
into account properly which is beyond the scope of the present paper.
\section{Conclusions}
In this paper we used the SCP and HzST data on high redshift type Ia supernovae and the 
radio source data from Gurvits et al. \cite{gurv} in order 
to test the infrared fixed point model proposed recently \cite{br2}. It predicts a time dependent 
cosmological and Newton constant whose dynamics arises from an underlying renormalization 
group flow, and it leads to a characteristic $t^{4/3}$-time dependence of the scale factor 
in the late Universe. The latter gives rise to a very particular distance-redshift 
relation which, in principle, can be verified or falsified using distant supernovae as 
standard candles and compact radio sources as standard rods.

The results of our investigation based upon the $\chi^2$ test can be summarized by saying 
that the IRFP model performs 
at least as well as the best-fit FRW model in reproducing the supernova and radio source data. 
Moreover, contrary to 
FRW cosmologies, it has {\it no free parameters} and thus actually {\it explains} why $\om$ 
and $\oa$ are of the same order today. In fact, in the flat case, they are predicted to 
be exactly equal ($\om=\oa=0.5$), today and at any time in the future.

In the space of all cosmologies with a power law expansion $a \propto t^\alpha$, 
it seems to be precisely the exponent $\alpha=4/3$ predicted by the IRFP model which yields 
the best fit to the data.

In this paper we also extended the original IRFP model of \cite{br2} by matching the fixed point 
regime with a preceeding FRW era. We found that the statistical quality of the currently 
available data is not sufficient in order to determine when the Universe entered the 
fixed point epoch. However, using a Monte Carlo simulation we saw that the data expected 
from the forthcoming SNAP mission can teach us something about the transition to the 
fixed point regime provided it happened late enough, at a redshift smaller than $0.5$, say.

In conclusion we think that the fixed point model is a viable alternative to the 
quintessence scenarios and certainly deserves being studied further. In particular it 
would be interesting to compare in more detail its predictions for the microwave background radiation 
to the data. We hope to address this issue in a future publication.

\section*{Acknowledgements}
We are very grateful to Ariel Goobar who has provided us with
the SNOC code for the Monte Carlo simulation. 
We would like to  thank V. Antonuccio, 
G. Esposito, G. Jorjadze, C. Kiefer, C. Rubano, G. Weigt and C. Wetterich 
for very interesting discussions. 
A.B. also acknowledges the warm 
hospitality of the Physics Departments of the University of Napoli and of 
Mainz University, where part of this work was written.
E.B. acknowledges the Astrophysical Observatory of Catania (OACt) for financial support.
M.R. is grateful to the OACt for the very 
kind hospitality extended to him. 

\appendix

\section{}


In this appendix we resume the analysis of the supernova data in terms of the IRFP model proper,
using a different statistical method: median statistics and a Bayesian model selection criterion.

\subsection{Median statistics}\label{sec:median}

Median statistics \cite{gott,avel} is more easily implemented than the $\chi^2$ analysis and also less 
demanding with respect to the 
assumptions about the data because uncertainties need not be normally distributed, 
and no prior knowledge of measurement errors is required.

According to median statistics, the operation of calculating the 
likelihood of a particular model boils down to counting: one simply enumerates
how many supernovae are, say, brighter than expected. Now, the probability that, 
{ when no systematic errors are present}, $n$ out of 
$92$ SNe are brighter than expected ($92-n$ being, of course, fainter) is given
by the standard binomial distribution $P(n,92)$.
As for the fixed point model proper (without a FRW epoch) and the data set of 
\ref{sec:data}, this number is found to be 
to $n=53$ which has binomial probability $P(53,92)=0.029$.
For comparison, we report in Table \ref{tab:table1} the same analysis for other
models. Notice that, if a model contains one or more free parameters, in 
column 1 we refer to those regions in the parameter space where the binomial 
probability is maximum.

\begin{table}[!h]
\bc
\caption{Binomial likelihood after 92 SNe. (For models with free parameters, the 
numbers pertain to those regions in parameter space where the probability is maximum.)
\label{tab:table1}}
\begin{ruledtabular}
\begin{tabular}{lccc}
\bf Model   &\bf Brighter        &\bf Fainter        &\bf Binomial likelihood
\\
\hline
Fixed point model ($z_{\rm tr}>z_{\rm max}$)&        $53$        &       $39$        &       $0.03$          
\\
Extended fixed point model ($z_{\rm tr} < 0.01$)	&	$46$	&	$46$	&	$0.08$\\

FRW with $\oaz=0$, $0\leq\omz\leq 0.1$   &  $40$        &       $52$        &         $0.04$      
\\
FRW with $\oaz+\omz=1$, $0.4\leq\omz\leq 0.5$&$46$        &       $46$        &         $0.08$      
\\
FRW with $\omz$ and $\oaz$ in the shaded region of Fig. \ref{fig:region}
&$46$&  $46$     &         $0.08$      
\\
\end{tabular}
\end{ruledtabular}
\ec
\end{table}

At first sight the IRFP model seems to perform rather poorly in comparison with the FRW 
models. However, it is important to understand that the figures in the last column of 
Table \ref{tab:table1} are not a measure for the correctness of the various models. In fact, 
the FRW models contain free parameters which always can be adjusted such that there 
is about the same number of brighter and fainter supernovae (at least if one allows 
for a cosmological constant). This is not possible with the IRFP model which contains 
no free parameters. Hence the naive median statistics is necessarily "unfair" towards 
the fixed point model. The correct way of assessing the "relative correctness" of the 
various models with their different number of free parameters is provided by Bayesian 
statistics, to which we shall turn in the next section.

\subsection{Bayesian model selection}
\subsubsection{The method}

In the Bayesian model selection the data set $DS$ is assumed to have arisen 
from one of several possible models (or hypotheses) $M_1, \cdots ,M_N$. 
An a {priori} probability
$P(M_i)$ is assigned to each model in order to  measure the likelihood of $M_i$ 
{when no other information} (e.g., observational) {is available}. 
This ``prior'' depends exclusively on the way each model is structured: we shall see 
below how to relate this number to the number of free parameters it presents. 
Let $P(DS|M_i)$ be the {likelihood} that a data set $DS$, such as the 
one that is actually observed, is attributable to $M_i$.
Then, following Bayes' theorem, the posterior probability $P(M_i|DS)$ that the model $M_i$ is responsible
for the observed data set $DS$ is
\be
\label{eq:final}
P(M_i|DS)= \frac{P(DS|M_i) \; P(M_i)}{\sum_{j} P(DS|M_j) \; P(M_j)}
\ee
The likelihood $P(DS|M_i)$ equals 
the statistical probability that, given $M_i$ is true, a set 
of $n$ observations would result in the measured outcome $DS$. Therefore, these numbers are 
just the experimentally determined binomial probabilities that we have already discussed 
in section \ref{sec:median}.

\subsubsection{Assigning {\it a priori} probabilities}
Let us now assume we are trying to determine the ``degree of belief'' 
$P(M_i)$ that model $M_i$ is the correct one, {\it before} we know anything about the
way it reproduces observations. Apart from the constraint that
\be
\label{eq:norm}
\sum_{i=1}^{N} P(M_i) = 1
\ee
we have no other clues. Basically, we would like to quantify the extent to 
which a given model is constructed on an {\it ad hoc} basis. According to 
Ockham's razor principle, the less free parameters are present,
the more a model can be regarded as physically plausible.
Bayesian priors can effectively help implementing this principle
if one relates the initial probabilities $P(M_i)$ of a model to the 
number of free parameters it contains. Following \cite{gott}, we then set
\be
P(M_i)=\frac {1} {2^{N_i+1}}
\ee
where $N_i$ is the number of free parameters of $M_i$.
It is easy to see that this definition guarantees (\ref{eq:norm}), provided 
we share the corresponding probability into equal parts whenever there is 
more than one model with the same $N_i$. 

\begin{figure}[ht]
\psfig{figure=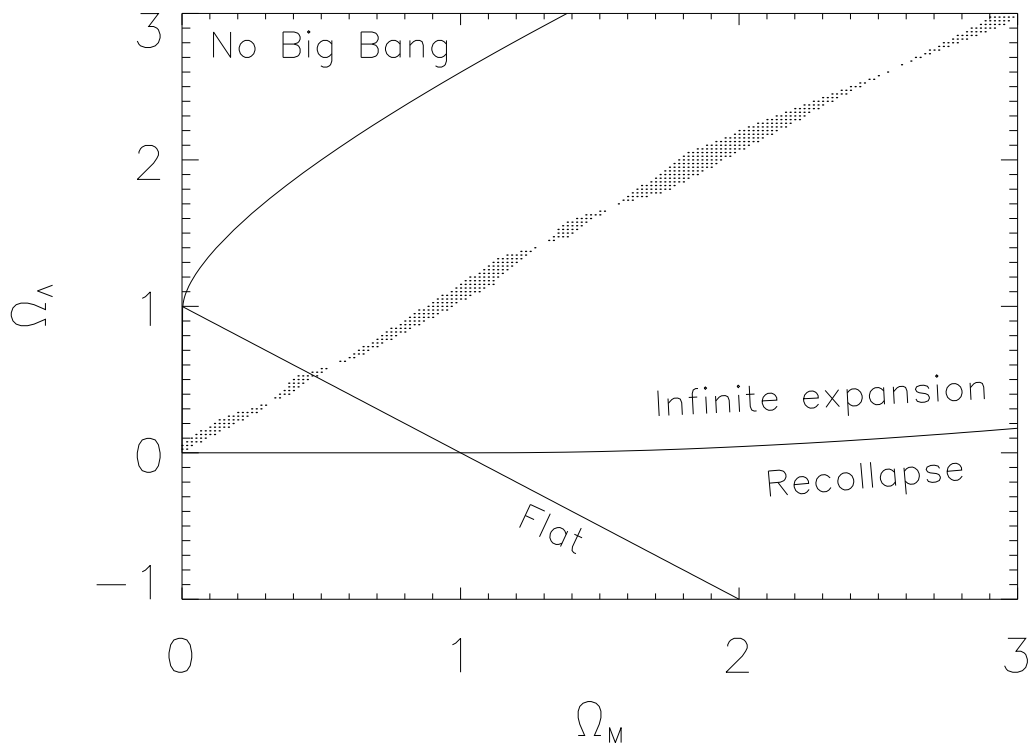,width=1.0\textwidth}
\caption{FRW cosmology: the values of ($\omz$, $\oaz$) in the shaded region 
maximize the binomial probability.
\label{fig:region}}
\end{figure}

In this paper we consider the following hypotheses:
\begin{enumerate}
\item $M_1$: The IRFP cosmology correctly describes the cosmic 
evolution for $z < z_{\rm tr}$, with $z_{\rm tr}>z_{\rm max}$;
\item $M_2$: The Universe is described by a zero-$\Lambda$ FRW cosmology, with 
$\omz > 0.1$;
\item $M_3$: The Universe is described by a flat FRW model, with $0.4 \leq \omz\leq 0.5$;
\item $M_4$: The Universe is described by a general FRW model, with $\omz$ and 
$\oaz$ both in the shaded confidence region indicated in Fig. \ref{fig:region}.
\end{enumerate}
The number of free parameters and the corresponding priors are listed in Table 
\ref{tab:table2}. (Here and in the following ``region 1'' stands for the shaded 
region in Fig. \ref{fig:region} where the binomial probability is maximum.)
The priors do not add up to unity yet; we shall comment on this point in a moment.

\begin{table}
\bc
\caption{Bayesian priors.\label{tab:table2}}
\begin{ruledtabular}
\begin{tabular}{lcc}
\bf Model                      &\bf Free parameters   &\bf Prior        \\
\hline
Fixed point model ($z_{\rm tr} > z_{\rm max}$)  &        none          &       $1/2$     \\
Extended fixed point model	&	1	&	1/12	\\
FRW with $\oaz=0$               &         1            &       $1/12$     \\
FRW with $\oaz+\omz=1$           &         1            &       $1/12$     \\
FRW with $\omz$ and $\oaz$ in region 1   &         2            &       $1/8$     \\
\end{tabular}
\end{ruledtabular}
\ec
\end{table}

\subsubsection{Comparison}
We are now in a position to evaluate the final probability to affix to each 
model $M_i$ as the ``degree of belief'' that, taking into account all the 
information in our possession, the hypothesis $M_i$ is true. All we have to 
do is to take each prior $P(M_i)$ listed in Table 
\ref{tab:table2}, multiply it by the measured binomial likelihood $P(DS|M_i)$ in 
Table \ref{tab:table1}, 
calculate the normalization factor ${\mathcal N} \equiv \sum_{i} P(DS|M_i)P(M_i)$,
and eventually compute the final probability $P(M_i|DS)$ according to (\ref{eq:final}). 

\begin{table}[!b]
\bc
\caption{Final probabilities.\label{tab:table3}}
\begin{ruledtabular}
\begin{tabular}{lc }
\bf Model                  &\bf Confidence model is true    \\
\bf                        &\bf (after 92 SNe Ia)           \\
\hline
Fixed point model  ($z_{\rm tr} > z_{\rm max}$)  &        $0.36$                    \\
Extended fixed point model	&	$0.16$		\\	
FRW with $\oaz=0$           &        $0.08$                    \\
FRW with $\oaz+\omz=1$       &        $0.16$                    \\
FRW with $\omz$ and $\oaz$ in region 1&    $0.24$              \\
\end{tabular}
\end{ruledtabular}
\ec
\end{table}

Our results are summarized in Table \ref{tab:table3}. The fixed point model is found 
to perform better than any standard cosmology. It has a posterior probability of $37.5 \%$, 
while the best-fit FRW model achieves only $25 \%$. Clearly these two numbers are 
not too different, but probably it is safe to say that the fixed point model is at 
least as likely to be the correct theory of the late Universe as is the best standard 
cosmology.
Also it should be kept in mind that the Bayesian result always depends on the choice 
for the priors. While our choice seems natural and has been used already in similar applications
\cite{gott} it is clear that others are conceivable as well.

Please note that the numbers in the last column of Table \ref{tab:table2} do not add 
up to unity yet. This is because we allow for models with more than 2 free parameters, 
whose prior probabilities, added up in the usual manner, $\sum_{N_i=3}^{+ \infty} P(M_i)$, 
cooperatively make up for the missing $1/8$.
However, we can safely ignore these cosmologies in the calculation of the final 
probabilities, because they are strongly penalized by their small priors, $1/2^{N_i+1}$ 
with $N_i>2$, and their inclusion in the computation does not significantly alter the 
results.

We can give a quantitative estimate of this assertion as follows: let us 
include in our discussion more and more cosmological models, $M_i$ ($i=5,\cdots$),  
each characterized by $N_i$ free parameters, with $N_i>2$. Their priors $P(M_i)$ are 
certainly such that
\be
\label{eq:prior}
P(M_i) \leq \frac{1}{16} \qquad \qquad (i>4)
\ee
On the other hand, their experimentally determined probabilities $P(DS|M_i)$ are subject to an 
upper bound, too, because they cannot exceed the maximum of the binomial likelihood 
function $P(n,92)$, which is obtained for $n=46$:
\be
\label{eq:exper}
P(DS|M_i) \leq P(46,92)=0.08
\ee
The increment in the normalization factor ${\mathcal N}=\sum_i P(DS|M_i)P(M_i)$, due to 
the new inclusions, is rather irrelevant:
\be
\delta {\mathcal N} = \sum_{i=6}^{+ \infty} P(DS|M_i)P(M_i) \leq 0.08 
\sum_{i=6}^{+ \infty} P(M_i) = 0.08 \sum_{N_i=3}^{+ \infty} \frac{1}
{2^{N_i+1}}=0.01
\ee
The value of ${\mathcal N}$, for the first 5 models only, amounts to $0.042$. Therefore, the effect of 
including all the possible cosmologies with $N_i>2$ is a shift of the value of 
${\mathcal N}$ from $0.042$ to (at most) $0.052$. 

By eq. (\ref{eq:final}), we also need
to recompute the final probabilities $P(M_i|DS)$. For the first 5 models, we have:
\be
\label{eq:new}
P^{\rm new}(M_i|DS)=\frac{P(DS|M_i)P(M_i)}{{\mathcal N}+ \delta {\mathcal N}}=
P^{\rm \, old}(M_i|DS) \frac{{\mathcal N}}{{\mathcal N}+ \delta {\mathcal N}}
\ee
Since
\be
\label{eq:deltaN}
0 \leq \delta {\mathcal N} \leq 0.01
\ee
the change in the final probabilities in Table \ref{tab:table3} 
reduces to a simple rescaling by a factor that satisfies
\be
0.81 \lesssim \frac{P^{\rm new}(M_i|DS)}{ P^{\rm \, old}(M_i|DS)} \leq 1
\ee
In particular, we observe that, irrespective of what the 
new models are
(as long as $N_i>2$, of course) or how well they compare to the data, the IRFP model will 
always have
\be
P^{\rm new}(M_1|DS) \gtrsim 29 \%
\ee

We might wonder how this figure compares to the new final probabilities for the other 
cosmologies. As for the four models with $N_i \leq 2$, since $P^{\rm \, old}(M_1|DS) > 
P^{\rm \, old}(M_i|DS)$, $i=2,3,4,5$, eq. (\ref{eq:new}) shows that
\be
P^{\rm new}(M_1|DS) > P^{\rm new}(M_i|DS)
\ee
in all cases. (The final probabilities are merely scaled, so their ratios remain unchanged.) 

As a last step, we must make sure that some of the new models do not acquire a final probability 
which is larger than $P^{\rm new}(M_1|DS)$. This possibility is easily ruled out by using 
(\ref{eq:prior}), (\ref{eq:exper}) and (\ref{eq:deltaN}):
\be
P^{\rm new}(M_i|DS)=\frac{P(DS|M_i)P(M_i)}{{\mathcal N}+ \delta {\mathcal N}} \leq
\frac{0.08 \cdot \frac{1}{16}}{0.052}=0.096 \qquad (i=6,\cdots)
\ee
Therefore, the new models cannot achieve a confidence level of more than $\sim 10\%$. 
This upper bound is based entirely on the fact that they contain a number of 
parameters which is larger than 2. Our final figures, reported in Table \ref{tab:table3}, 
reflect precisely these  considerations, which fully entitle us to ignore any 
model with $N_i>2$.



\begin{thebibliography}{9999999}
\bibitem{classav} M. Carfora, K. Piotrkowska, Phys.Rev. D53 (1995) 4393;
T. Buchert, M. Carfora, \\ 
gr-qc/0101070;  T. Buchert, M. Carfora, gr-qc/0210045;
\bibitem{bel} J.A. Belinchon, T. Harko, M.K. Mak, Class.Quant.Grav. 19 (2002) 3003 
and \\
gr-qc/0108074, and references therein.
\bibitem{ber}For a review see: J. Berges, N. Tetradis, C. Wetterich, hep-ph/0005122;\\
C. Wetterich, hep-ph/0101178.
\bibitem{mr}M. Reuter, Phys.Rev. D57 (1998) 971 and hep-th/9605030. 
\bibitem{geo}For a brief introduction see:
M. Reuter in {\it Annual Report 2000 of the International School in Physics and
Mathematics, Tbilisi, Georgia} and hep-th/0012069.
\bibitem{ol} O. Lauscher, M. Reuter, Phys.Rev. D65 (2001) 025013 
and hep-th/0108040.
\bibitem{ol2} O. Lauscher, M. Reuter, Class.Quant.Grav. 19 (2002) 483 and
hep-th/0110021; 
Phys.Rev. D66 (2002) 025026 and hep-th/0205062; 
Int.J.Mod.Phys. A17 (2002) 993 and hep-th/0112089.
\bibitem{frank} M. Reuter, F. Saueressig,
Phys.Rev. D65 (2002) 065016 and hep-th/0110054;
Phys.Rev. D66 (2002) 125001 and hep-th/0206145.
\bibitem{per}R. Percacci, D. Perini, hep-th/0207033; D. Dou, R. Percacci, Class.Quant.Grav. 15 (1998) 3449.
\bibitem{souma} W. Souma, Prog.Theor.Phys. 102 (1999) 181.
\bibitem{max} For a recent discussion of the UV fixed point within the 2 Killing vector reduction of QEG see:
P. Forg\'acs and M. Niedermaier, hep-th/0207028;
M. Niedermaier, hep-th/0207143.
\bibitem{br1}A. Bonanno, M. Reuter, Phys.Rev. D65 (2002) 043508
and hep-th/0106133.
\bibitem{cw}M. Reuter, C. Wetterich, Phys.Lett. B188 (1987) 38.
\bibitem{bh2}A. Bonanno, M. Reuter, Phys.Rev. D62 (2000) 043008 and hep-th/0002196.
\bibitem{bh1}A. Bonanno, M. Reuter, Phys.Rev. D60 (1999) 084011 and gr-qc/9811026.
\bibitem{br2}A. Bonanno, M. Reuter, Phys.Lett. B527 (2002) 9
and astro-ph/0106468.
\bibitem{coscon2} V. Sahni, A. Starobinsky, astro-ph/9904398;
N. Straumann, astro-ph/9908342.
\bibitem{tsamis}For earlier work in this direction, see \\
N.C. Tsamis, R.P. Woodard, Phys.Lett. B301 (1993) 351; Ann.Phys. (NY) 238 (1995) 1; \\
I. Antoniadis, E. Mottola, Phys.Rev. D45 (1992) 2013;\\
I. Antoniadis, P. Mazur, E. Mottola, Phys.Lett. B444 (1998) 284.
\bibitem{pert} A. Bonanno, M. Reuter, astro-ph/0210472, to appear on IJMPD.
\bibitem{gott} J.R. Gott et al., Ap.J. 549 (2001) 1.
\bibitem{avel} P.P. Avelino et al., Astrophys.J. 575 (2002) 989 
\bibitem{perl}S. Perlmutter et al., Astrophys.J. 517 (1999) 565.
\bibitem{riess}A. Riess et al., Astron.J. 117 (1999) 707.
\bibitem{berto}For a similar cosmology in a Brans-Dicke framework see\\
O. Bertolami, P.J. Martins, Phys.Rev. D61 (2000) 064007.
\bibitem{liouv} M. Reuter, C. Wetterich, Nucl.Phys. B506 (1997) 483;\\
A. Chamseddine, M. Reuter, Nucl.Phys. B317 (1989) 757.
\bibitem{hamuy}Hamuy et al., Astrophys.J. 109 (1995) 1.
\bibitem{goobar} A. Goobar, E. M\"ortsell, R. Amanullah, M. Goliath, L. Bergstr\"om, T. Dahlen,
Astron. \& Astrophys. 392 (2002) 757.
\bibitem{gurv}L.I. Gurvits, K.I. Kellermann, S. Frey, Astron. Astrophys. 342 (1999) 378.
\bibitem{jain} D. Jain, A. Dev, J.S. Alcaniz, astro-ph/0302025.
\bibitem{zhu} Z. Zhu, M. Fujimoto, Ap.J. 581 (2002) 1.
\bibitem{sys1}
D. Kalligas, P. Wesson, C.W.F. Everitt, Gen.Rel.Grav. 24 (1992) 351. See also:\\
A. Beesham, Nuovo Cimento 96B (1986) 17, Int.J.Theor.Phys. 25 (1986) 1295;\\
A.-M.M. Abdel-Rahman, Gen.Rel.Grav. 22 (1990) 655;\\
M.S. Berman, Phys.Rev. D43 (1991) 1075, Gen.Rel.Grav. 23 (1991) 465;\\
R.F. Sistero, Gen.Rel.Grav. 23 (1991) 1265;\\
T. Singh, A. Beesham, Gen.Rel.Grav. 32 (2000) 607;\\
A. Arbab, A. Beesham, Gen.Rel.Grav. 32 (2000) 615;
A. Arbab, gr-qc/9909044.
\bibitem{wmap}C. Bennet et al., Ap.J., in press, and astro-ph/0302207.
\bibitem{weinp}S. Weinberg, Phys.Rev. D64 (2001) 123511.
\bibitem{angle}L. Knox, N. Christensen, C. Skordis, Ap.J. 581 (2001) L95.

\end{thebibliography}
\end{document}